\documentclass[
	reprint,
	amsmath,amssymb,
	superscriptaddress,
	twocolumn,
	accepted=2023-04-04
   ]{quantumarticle}
\usepackage[numbers,sort&compress,square]{natbib}

\usepackage{graphicx}
\usepackage{color}
\usepackage{dcolumn}
\usepackage{siunitx}
\usepackage{bm}
\usepackage{braket}
\usepackage{tabularx}
\usepackage{longtable}
\newtheorem{theorem}{Theorem}
\begin{document}


\title[]{Perturbation theory with quantum signal processing}

\author{Kosuke Mitarai}
\email{mitarai.kosuke.es@osaka-u.ac.jp}
\affiliation{Graduate School of Engineering Science, Osaka University, 1-3 Machikaneyama, Toyonaka, Osaka 560-8531, Japan.}
\affiliation{Center for Quantum Information and Quantum Biology, Osaka University, Japan.}

\author{Kiichiro Toyoizumi}
\affiliation{Graduate School of Science and Technology, Keio University, 3-14-1 Hiyoshi, Kohoku-ku, Yokohama 223-8522, Japan.}

\author{Wataru Mizukami}
\affiliation{Center for Quantum Information and Quantum Biology, Osaka University, Japan.}

\date{\today}

\begin{abstract}
    Perturbation theory is an important technique for reducing computational cost and providing physical insights in simulating quantum systems with classical computers.
	Here, we provide a quantum algorithm to obtain perturbative energies on quantum computers. 
	The benefit of using quantum computers is that we can start the perturbation from a Hamiltonian that is classically hard to solve.
	The proposed algorithm uses quantum signal processing (QSP) to achieve this goal.
	Along with the perturbation theory, we construct a technique for ground state preparation with detailed computational cost analysis, which can be of independent interest.
	We also estimate a rough computational cost of the algorithm for simple chemical systems such as water clusters and polyacene molecules.
	To the best of our knowledge, this is the first of such estimates for practical applications of QSP.
	Unfortunately, we find that the proposed algorithm, at least in its current form, does not exhibit practical numbers despite of the efficiency of QSP compared to conventional quantum algorithms.
	However, perturbation theory itself is an attractive direction to explore because of its physical interpretability; it provides us insights about what interaction gives an important contribution to the properties of systems.
	This is in sharp contrast to the conventional approaches based on the quantum phase estimation algorithm, where we can only obtain values of energy.
	From this aspect, this work is a first step towards ``explainable'' quantum simulation on fault-tolerant quantum computers. 
\end{abstract}

\maketitle

\section{Introduction}

Perturbation theory is one of the most important techniques to understand quantum systems.
It solves problems by separating them into easy parts and difficult ones, and gradually taking the effect of difficult parts into account.
For weakly correlated systems, it usually gives sufficiently accurate physics of the systems.
A benefit of using perturbative methods is its computational efficiency compared to the cost of exactly solving quantum systems.
The computational cost to obtain an exact solution of an $n$-body quantum system is generally exponential to $n$ on classical computers, while that of perturbative methods is only polynomial.
Another important aspect of perturbation is its physical interpretability.
It provides us insights into what effect a specific interaction of the system has on its overall physical properties.

In this work, we provide a method to implement perturbation theory on quantum computers and discuss if the benefits described above can also be obtained.
Our method allows one to use strongly-interacting Hamiltonians that are only solvable with quantum computers as a starting point of the perturbation.
More specifically, our algorithm first constructs a ground state of an unperturbed Hamiltonian via quantum signal processing (QSP) \cite{Low2017QSP, GrandUnification, QSVT} and fixed-point amplitude amplification \cite{FixedPointSearch2014}.
Then, we generate a perturbative state by applying the inverse of an unperturbed Hamiltonian with quantum signal processing (QSP), and obtain an expectation value of a perturbation operator via robust amplitude estimation (RAE) \cite{RAE-first, RAE-VQE, RAEexperiment}.

For an unperturbed Hamiltonian $H$ and perturbation $V$ that can be written as $H=\sum_{\ell} h_\ell \sigma_\ell$ and $V=\sum_{\ell} v_\ell \sigma_\ell$ where $\sigma_\ell$ are Pauli operators, 
the complexity of the proposed algorithm is $\tilde{\mathcal{O}}(\|\bm{h}\|_1\|\bm{v}\|_{2/3}/(\Delta \delta))$ and $\tilde{\mathcal{O}}(\|\bm{h}\|_1\|\bm{v}\|_{2/3}^2/(\Delta^2 \delta))$ respectively for the first-order and second-order perturbation, where $\Delta$ is a spectral gap of $H$, $\delta$ is error in the estimated perturbation energy and $\|\bm{a}\|_{p} = (\sum_\ell a^{p}_\ell)^{1/p}$.

We also perform a concrete resource analysis of the algorithm for simple chemical systems such as water clusters and polyacenes to discuss its practicality.
This is, to the best of our knowledge, the first such analysis of a practical application of the QSP and the QSP-based matrix inversion technique.
Despite of efficiency of QSP compared to conventional techniques, it is found that our algorithm gives impractical numbers as computational cost; for example, we estimate over $10^{31}$ calls of block-encodings would be required to perform perturbation on a pentacene molecule.
This is much larger than the cost required for naively performing the phase estimation of the total Hamiltonian, which only requires $10^{10}$ calls of block-encodings.
While we could not achieve a reduction of computational cost like the classical perturbation theory, the other benefit of perturbation, that is, the interpretability of the result, is still an important point.
Conventional techniques of quantum simulations based on phase estimation can give us energy and its eigenstates, but cannot provide us insights into why the energy is the obtained value.
We, therefore, believe this work is a first step toward an ``explainable'' quantum simulation on fault-tolerant quantum computers.

\section{Preliminaries}

\subsection{Perturbation theory}

We have an $n$-qubit Hamiltonian $H_{\mathrm{total}}= H+V$.
We consider the Hamiltonian $H$ and the perturbation $V$ that can be decomposed into Pauli operators $\sigma_{\ell}$ as,
\begin{align}
    H = \sum_{\ell=1}^{L_H} h_\ell \sigma_\ell \\
    V = \sum_{\ell=1}^{L_V} v_\ell \sigma_\ell \label{eq:V-def}
\end{align}
Note that $H$ and $V$ can, without loss of generality, be defined as not having $I^{\otimes n}$ term.
Let the ground state of $H_{\mathrm{total}}$ with eigenvalue $E_0$ be $\ket{E_0}$.
Also, let an eigenstate $H$ with an eigenvalue $\epsilon_i$ be $\ket{\epsilon_i}$.
We assume that the eigenvalues are ordered in ascending order $\epsilon_{0}< \epsilon_{1} \leq \cdots \leq \epsilon_{2^n-1}$.

It is well known \cite{doi:https://doi.org/10.1002/9781119019572.ch14} that $\ket{E_0}$ can be approximated as
\begin{align}
    \ket{E_0} &\approx \ket{\epsilon_0^{(1)}} \\
    & := \ket{\epsilon_0} - \Pi (H-\epsilon_0)^{-1}\Pi V\ket{\epsilon_0}, \label{eq:perturbed-state}
\end{align}
where $\Pi=I-\ket{\epsilon_0}\bra{\epsilon_0}$,
to the first order in $\|V\|$ if $\epsilon_i$ is not degenerate.
The corresponding eigenvalue $E_0$ can be approximated as,
\begin{align}
    E_0 \approx \epsilon_0 + \epsilon_0^{(1)} + \epsilon_0^{(2)}, \label{eq:perturbative-energy}
\end{align}
where
\begin{align}\label{eq:first-order-perturbation}
    \epsilon_0^{(1)} = \braket{\epsilon_0|V|\epsilon_0}
\end{align}
and
\begin{align}\label{eq:second-order-perturbation}
    \epsilon_0^{(2)} = -\bra{\epsilon_0}V \Pi (H-\epsilon_0)^{-1}\Pi V \ket{\epsilon_0}.
\end{align}

\subsection{Block-encoding of a Hamiltonian}\label{sec:block-encoding}

First, we introduce the block-encoding \cite{chakraborty_et_al:LIPIcs:2019:10609, QSVT}.
We say a unitary $U_A$ block-encodes a matrix $A$ when it has the following form:
\begin{equation}
    U_A = \left(
        \begin{array}{cc}
            A/\alpha & \cdot \\
            \cdot & \cdot
        \end{array}
    \right).
\end{equation}
More formally, we say $(n+l)$-qubit unitary $U_A$ is a $(\alpha, l, \varepsilon)$-block encoding of $n$-qubit operator $A$ if
\begin{equation}
    \left\|A-\alpha\left(\bra{0^l} \otimes I\right) U\left(\ket{0^l}\otimes I\right) \right\| \leq \varepsilon.
\end{equation}
Babbush \textit{et al.} \cite{PhysRevX.8.041015} showed that we can perform phase estimation of a unitary $e^{i \arccos A/\alpha}$ to estimate an eigenvalue of $A$ to precision $\delta$ by using $U_A$ for $\frac{\sqrt{2}\pi \alpha}{2\delta}$ times.
The phase estimation algorithm \cite{Nielsen2010-ex} takes in a state $\ket{\psi}$ and outputs eigenvalue $e^{i\phi}$ of a unitary $U$ with a probability $p(\phi)=|\braket{\phi|\psi}|^2$ where $\ket{\phi}$ is the eigenstate of $U$ corresponding to the eigenvalue $e^{i\phi}$. 
If we wish to obtain a specific eigenvalue such as the ground state energy, $\ket{\psi}$ must therefore have non-negligible overlap with the corresponding eigenstate.

An explicit block-encoding of an $n$-qubit operator $A$ which can be represented as a sum of Pauli operator as $A=\sum_{\ell=1}^L a_\ell \sigma_\ell$ can be constructed via $\texttt{PREPARE}_{\bm{a}}$ and \texttt{SELECT} operations introduced in Ref. \cite{PhysRevX.8.041015}.
$\texttt{PREPARE}$ operation acts on $l = \lceil\log_2 L\rceil$ qubits as 
\begin{equation}
    \texttt{PREPARE}_{\bm{a}}\ket{0^l} =\sum_{\ell=0}^{L-1} \sqrt{\frac{a_{\ell}}{\|\bm{a}\|_1}}\ket{\ell} \equiv \ket{\mathcal{L}_\mu}.
\end{equation}
where $\|\bm{a}\|_1 =\sum_\ell |a_{\ell}|$.
\texttt{SELECT} acts on $n+l$ qubits and defined as 
\begin{equation}
    \texttt{SELECT} = \sum_{\ell=0}^{L-1}\ket{\ell}\bra{\ell}\otimes \sigma_\ell.
\end{equation}
Then,
\begin{align}
    U_A = \texttt{PREPARE}^\dagger_{\bm{a}} \cdot \texttt{SELECT} \cdot \texttt{PREPARE}_{\bm{a}}
\end{align}
satisfies $(\bra{0^l}\otimes I_n) U_A(\ket{0^l}\otimes I_n) = A/\|\bm{a}\|_1$, i.e., $U_A$ is a $(\|\bm{a}\|_1,l,0)$-block-encoding of $A$.
It has been shown that $\texttt{PREPARE}_{\bm{a}}$ and \texttt{SELECT} operations can be implemented with $O(L+\log(1/\varepsilon))$ T gates.
Note that the T gate is the most time-consuming gate in the surface-code-based fault-tolerant quantum computing \cite{PhysRevX.8.041015}.

\subsection{Eigenvalue transformation via quantum signal processing}\label{sec:QSP}
Given a block-encoding $U_A$ of $A$, we can construct a block encoding of $P(A)$ for certain polynomials $P(x)$ \cite{GrandUnification,QSVT,Low2017QSP}.
In this work, we are only interested in real polynomials $P(x)$.
The seminal work \cite[Corollary 10]{QSVT} shows that, for a degree-$d$ real polynomial $P(x)$ such that
\begin{align}
    &\text{$|P(x)|\leq 1$ for $|x|\leq1$, } \label{eq:QSP-condition1}\\
    \begin{split}
        &\text{$P(x)=P(-x)$ if $d$ is even,} \\
        &\text{$P(x)=-P(-x)$ if $d$ is odd}
    \end{split}\label{eq:QSP-condition2}
\end{align}
we can obtain a unitary $\mathcal{P}_0$ and $\mathcal{P}_1$ such that
\begin{align}
    \mathcal{P}_0 \ket{0^l}\ket{\psi} &= \ket{0^l}[\tilde{P}(A)\ket{\psi}] + \ket{g},\\
    \mathcal{P}_1 \ket{0^l}\ket{\psi} &= \ket{0^l}[\tilde{P}^*(A)\ket{\psi}] + \ket{g},
\end{align}
where $\tilde{P}(x)$ is a degree-$d$ polynomial such that $\mathrm{Re}[\tilde{P}(x)]=P(x)$, $^*$ denotes complex conjugate, and $\ket{g}$ is a ``garbage'' state which is orthogonal to $\ket{0^l}[\tilde{P}(A)\ket{\psi}]$.
We can construct $\mathcal{P}_0$ and $\mathcal{P}_1$ with $d$ calls of $U_A$.
Moreover, a unitary defined as
\begin{align}
    \mathcal{P} = \ket{+}\bra{+}\otimes \mathcal{P}_0 + \ket{-}\bra{-}\otimes \mathcal{P}_1,
\end{align}
which uses another ancilla qubit, satisfies
\begin{align}
    \mathcal{P}\ket{0^{l+1}}\ket{\psi} &= \ket{0^{l+1}}[P(A)\ket{\psi}] + \ket{g}
\end{align}
as shown in \cite[Corollary 18]{QSVT}.
Note that $\mathcal{P}$ can be constructed by only using $d$-calls of $U_A$ rather than $2d$-calls, since $\mathcal{P}_1$ is obtained by inverting phase sequences of $\mathcal{P}_0$ \cite[Corollary 18]{QSVT}. 
After the application of $\mathcal{P}$, we can post-select on the ancilla being $\ket{0^{l+1}}$ to obtain a state proportional to $P(A)\ket{\psi}$.
The probability of success is $\|P(A)\ket{\psi}\|^2$.
This procedure is called quantum signal processing (QSP) \cite{GrandUnification, Low2017QSP, QSVT}.
In this work, we say a polynomial is QSP-implementable if it satisfies the above two conditions \eqref{eq:QSP-condition1} and \eqref{eq:QSP-condition2}.

\subsection{Amplitude estimation}

Amplitude estimation refers to various techniques to obtain values of amplitudes of a quantum state $\ket{\psi}$ within error of $\delta$ using $\mathcal{O}(1/\delta)$ calls of a state preparation unitary $U_\psi$ such that $\ket{\psi}= U_\psi\ket{0}$.
The original algorithm can be found in \cite{quant-ph/0005055v1}.
Recent developments have made the procedure significantly more efficient.
For example, a state-of-the-art method called the robust amplitude estimation (RAE) \cite{RAE-first, RAE-VQE, RAEexperiment} can empirically estimate $\braket{\psi|\sigma|\psi}$ for a Pauli operator $\sigma$ within a mean squared error of $\delta^2$ by using 
\begin{align}
    \frac{5\sqrt{2}}{2}\frac{e^2}{e-1}\frac{1}{\delta}
\end{align}
calls of $U_\psi$.
Other recent works such as iterative quantum amplitude estimation \cite{Grinko_2021,2207.08628v2} have shown similar performance. 
In this work, we employ RAE to estimate the expectation values.

\section{Perturbation theory on quantum computers}

\subsection{High-level overview}
Our approach for performing perturbation on a quantum computer is as follows:
\begin{enumerate}
    \item Perform phase estimation of $H$ to estimate  $\epsilon_0$ within a precision of $\delta_0$.
    \item Efficiently generate $\ket{\epsilon_0}$ via QSP-based eigenstate filtering (Sec. \ref{sec:ground-state-preparation}).
    \item Estimate the first-order perturbation energy by measuring $\braket{\epsilon_0|V|\epsilon_0}$ (Sec. \ref{sec:first-order}).
    \item Estimate the second-order perturbation energy by performing the Hadamard test of a unitary which approximates $\Pi (H-\epsilon_0)^{-1}\Pi$ constructed via QSP (Sec. \ref{sec:second-order}). 
\end{enumerate}
Step 1 of the algorithm can be replaced with more sophisticated techniques provided in Refs.~\cite{Lin2020nearoptimalground, PRXQuantum.3.010318} from the naive phase estimation.
Algorithms for step 2 are also proposed in e.g. \cite{Ge2017filtering,Lin2020nearoptimalground,FTQC-derivative}, but previous works discuss the cost in terms of $\mathcal{O}$-notations.
In the following subsections, we describe the details of steps 2, 3, and 4, without resorting to $\mathcal{O}$-notations.
Some assumptions we make are in order.
First, we assume that one can prepare a state $\ket{\psi}$ that has non-zero overlap $p$ with $\ket{\epsilon_0}$ in steps 1 and 2.
We do not discuss how to choose and prepare such a state in detail because they strongly depend on the target system.
For our numerical resource estimation in Sec. \ref{sec:resource-analysis}, we employ Hartree-Fock states.
The second assumption is the knowledge of lower bounds to the overlap $p$ and the spectral gap $\Delta=\epsilon_1-\epsilon_0$ of the unperturbed Hamiltonian $H$.
This assumption is required for constructing quantum circuits for steps 2 and 4.
In the following, we formulate the cost by using $p$ and $\Delta$ but they can always be replaced by their lower bounds.

We note that Ref. \cite{GreenFunctionQSP} has considered a similar task that involves the inversion of Hamiltonians using QSP.
Their main target is, however, to calculate Green's function, which is widely used to obtain response properties with respect to external fields, and not to construct general perturbation theory, which can evaluate the energy of a many-body system, like this work.
From the technical viewpoint, for the calculation of Green's function, only the operator in the form of $(H-z)^{-1}$ where $z$ is a complex number is required and one does not need to construct $\Pi (H-\epsilon_0)^{-1}\Pi$ as we do here.
Ref. \cite{endosan} also considers the perturbative simulation of quantum systems.
Their goal is to implement real-time dynamics on a quantum computer and does not focus on obtaining the energies of Hamiltonians.
It is achieved by first discretizing the time evolution to small time slices and sampling the perturbative part of each of the short-time dynamics with respect to a quasi-probability distribution.
Their method hence requires exponential cost with respect to the evolution time, and would not be suitable for extracting energy eigenvalues with high precision.

\subsection{Preparation of a reference state}\label{sec:ground-state-preparation}
\begin{figure}
    \centering
    \includegraphics[width=\linewidth]{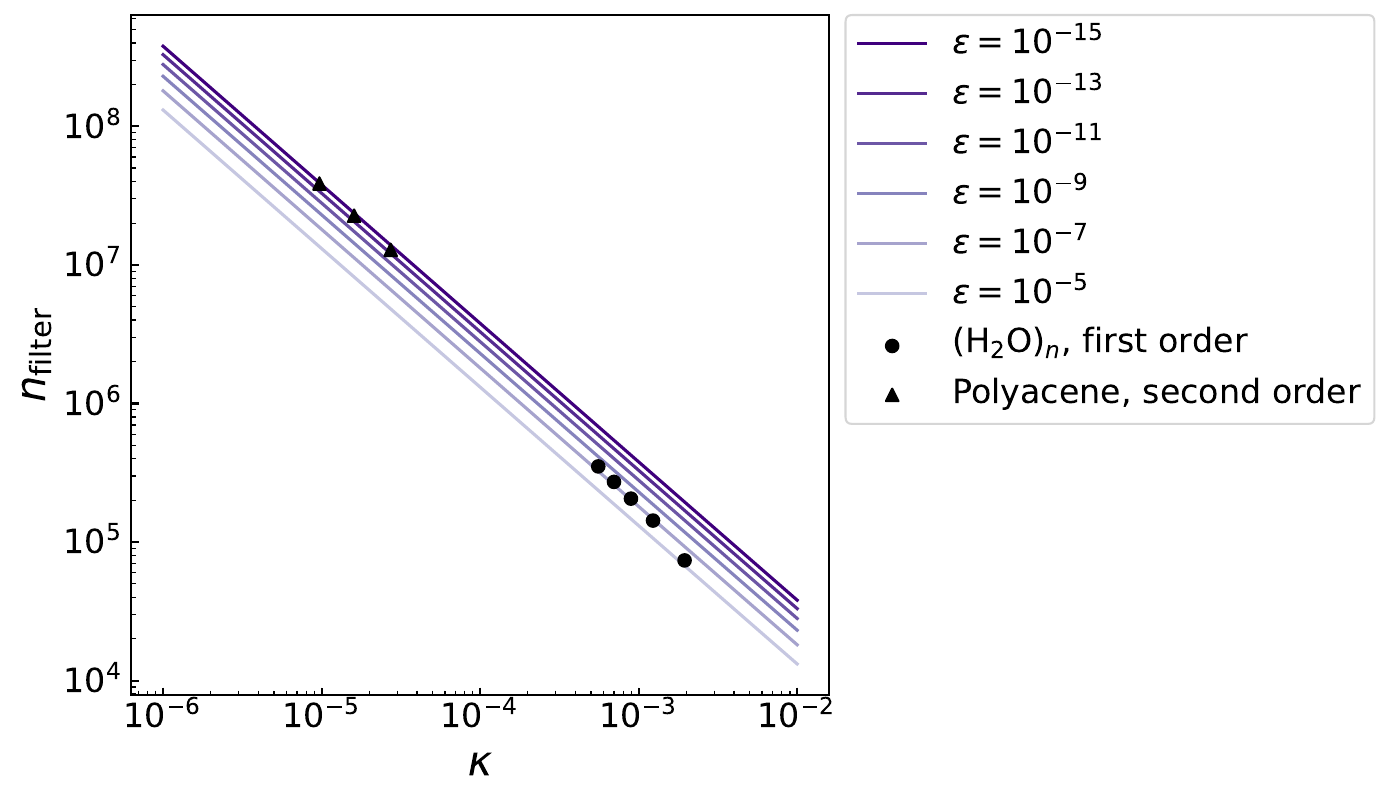}
    \caption{\label{fig:filtering} Values of $n_{\mathrm{filter}}$ calculated by Eq.~\eqref{eq:filtering-degree} with $x_{\mathrm{th}}=10^{-6}$, which is a typical value for the molecules studied in Sec.~\ref{sec:resource-analysis}, and with different error parameters $\varepsilon$ as a function of $\kappa$. Points corresponds to the values for specific molecules presented in Tables \ref{tab:first_order_perturbation} and \ref{tab:second_order_perturbation}.}
\end{figure}

In this work, we utilize the QSP-approximation of rectangular functions introduced by Low and Chuang \cite{Low2017uniformspectral} and reviewed in Ref. \cite{GrandUnification} to prepare the reference state $\ket{\epsilon_0}$ from which we start the perturbation.
$\ket{\epsilon_0}$ can be generated via phase estimation.
However, it is more efficient to use QSP to filter $\ket{\epsilon_0}$ when we know the value of the corresponding eigenvalue $\epsilon_0$.
The construction of rectangular functions closely follows that of \cite{Low2017uniformspectral, GrandUnification} but we give more detailed cost, that is, the degree of polynomial needed for their approximation, than the previous works.
In Appendix \ref{appsec:step-polynomial}, we show that there exists a QSP-implementable polynomial $P^{\mathrm{filter}}_{\varepsilon,\kappa,x_{\mathrm{th}}}(x)$ such that,
\begin{align}\label{eq:filtering-property}
    \begin{split}
        P^{\mathrm{filter}}_{\varepsilon,\kappa,x_{\mathrm{th}}}(x) &> 1-\varepsilon~~(|x|<x_{\mathrm{th}}), \\
    |P^{\mathrm{filter}}_{\varepsilon,\kappa,x_{\mathrm{th}}}(x)| &< \varepsilon~~(|x|>x_{\mathrm{th}}+\kappa),
    \end{split}
\end{align}
where $x_{\mathrm{th}}>0$, $0<\kappa<2(1-x_{\mathrm{th}})$ are parameters, with degree roughly
\begin{align}\label{eq:filtering-degree}
    \begin{split}
        &n_{\mathrm{filter}}(\varepsilon, \kappa, x_{\mathrm{th}}) \\
        &\approx \frac{64 (1+x_{\mathrm{th}}')}{\sqrt{\pi} \varepsilon} \frac{1}{\kappa}\sqrt{2\log_e \left(\frac{8}{\pi\varepsilon^2}\right)} \exp\left(- \frac{1}{2}W\left(\frac{2048}{\pi \varepsilon^{2} e^{2}}\right)\right), 
    \end{split}
\end{align}
where $x_{\mathrm{th}}'=x_{\mathrm{th}}+\kappa/2$ and $W(x)$ is the Lambert W function.
We plot the values of $n_{\mathrm{filter}}(\varepsilon, \kappa, x_{\mathrm{th}})$ as Fig.~\ref{fig:filtering} with various parameter settings.
Using an inequality $ W(x) > \log(x) - \log(\log(x))$, we can roughly upper-bound $n_{\mathrm{filter}}(\varepsilon, \kappa, x_{\mathrm{th}})$ by,
\begin{align}\label{eq:filtering-degree-approx-upper-bound}
    \begin{split}
        &n_{\mathrm{filter}}(\varepsilon, \kappa, x_{\mathrm{th}}) \\
        &\leq \frac{e (1+x_{\mathrm{th}}')}{\sqrt{2}\kappa} \sqrt{\log_e \left(\frac{8}{\pi\varepsilon^2}\right)\log_e\left(\frac{2048}{\pi \varepsilon^{2} e^{2}}\right)}, 
    \end{split}
\end{align}
which has $\mathcal{O}(\log(1/\varepsilon)/\kappa)$ scaling.

Let us assume that we know an estimate $\hat{\epsilon}_0$ of $\epsilon_0$ such that $|\hat{\epsilon}_0-\epsilon_0|<\delta_0$.
Define $H'=H-\hat{\epsilon}_0I$.
Let
\begin{align}
    H' = \sum_{\ell=1}^{L_H+1} h_\ell' \sigma_\ell.
\end{align}
Note that, since we assume that $H$ does not have $I^{\otimes n}$ term, $H'$ has $L_H+1$ terms.
Also, note that $\|\bm{h}'\|_1 = \|\bm{h}\|_1 + |\hat{\epsilon}_0|$.
Let $l'=\lceil\log(L_H+1)\rceil$, the number of ancilla qubits needed to block-encode $H'$.
We can perform QSP of $H'$ via $U_{H'}$.
QSP can construct $P^{\mathrm{filter}}_{\varepsilon,\kappa}(H'/\|\bm{h}'\|_1)$ with $n_{\mathrm{filter}}(\varepsilon, \kappa, x_{\mathrm{th}})$ calls of $U_{H'}$ (see Sec. \ref{sec:QSP}).
Then, since we know the ground state energy of $H'/\|\bm{h}'\|_1$ is within $[-\delta_0/\|\bm{h}'\|_1,\delta_0/\|\bm{h}'\|_1]$ and the second largest energy is larger than $(\Delta-\delta_0)/\|\bm{h}'\|_1$, we can set
\begin{align}
    x_{\mathrm{th}} &= \delta_0/\|\bm{h}'\|_1, \label{eq:x_th}\\
    \kappa &= (\Delta-\delta_0)/\|\bm{h}'\|_1, \label{eq:kappa}
\end{align}

The error parameter $\varepsilon$ controls the fidelity of $\ket{\epsilon_0}$.
Let us assume an initial state $\ket{\psi}$ fed to the QSP satisfies,
\begin{align}\label{eq:initial-state}
    \ket{\psi} = \sqrt{p}\ket{\epsilon_0} + \sqrt{1-p}\ket{\epsilon_0^{\perp}},
\end{align}
where $\ket{\epsilon_0^{\perp}}$ is a state orthogonal to $\ket{\epsilon_0}$.
To obtain $\ket{\epsilon_0}$, we first apply $P^{\mathrm{filter}}_{\varepsilon,\kappa,x_{\mathrm{th}}}(H'/\|\bm{h}'\|_1)$ via QSP and get a state in the form of,
\begin{align}
    \ket{0^{l'+1}}\left[P^{\mathrm{filter}}_{\varepsilon,\kappa,x_{\mathrm{th}}}\left(\frac{H'}{\|\bm{h}'\|_1}\right)\ket{\psi}\right] + \ket{g}.
\end{align}
The post-selection on the ancilla being $\ket{0}$ results in a state 
\begin{align}
    \ket{\tilde{\epsilon}_0} = P^{\mathrm{filter}}_{\varepsilon,\kappa,x_{\mathrm{th}}}(H'/\|\bm{h}'\|_1)\ket{\psi}/\|P^{\mathrm{filter}}_{\varepsilon,\kappa,x_{\mathrm{th}}}(H'/\|\bm{h}'\|_1)\ket{\psi}\|, \label{eq:approx-ground-state}
\end{align}
which is used as the reference state for the perturbation.
We obtain the perturbation energies as expectation values of observables $O=V$ and $V\Pi(H-\epsilon_0 )\Pi V$ with respect to $\ket{\tilde{\epsilon}_0}$.
We, therefore, wish to make 
\begin{align}
    \delta_{\mathrm{prep}} := |\braket{\epsilon_0|O|\epsilon_0} - \braket{\tilde{\epsilon}_0|O|\tilde{\epsilon}_0}| \ll \delta
\end{align}
to obtain the perturbation energy with an accuracy of $\delta$.
In Appendix \ref{appsec:prep-error}, it is shown that,
\begin{align}
    \delta_{\mathrm{prep}} = 2\varepsilon\sqrt{\frac{1-p}{p}}\mathrm{Re}\braket{\epsilon_0^{\perp}|O|\epsilon_0} + \mathcal{O}(\varepsilon^2) \label{eq:delta_prep}
\end{align}
Assuming $\mathcal{O}(\varepsilon^2)$ term is negligible, we can roughly upper-bound $\delta_{\mathrm{prep}}$ for $O=V$ and $V\Pi(H-\epsilon_0 )\Pi V$ respectively as,
\begin{align}\label{eq:delta_prep_first_order}
    \delta_{\mathrm{prep}} \lesssim 2\varepsilon\sqrt{\frac{1-p}{p}}\|V\|
\end{align}
and 
\begin{align}\label{eq:delta_prep_second_order}
    \delta_{\mathrm{prep}} \lesssim 2\varepsilon\sqrt{\frac{1-p}{p}}\frac{\|V\|^2}{\Delta} 
\end{align}

We can remove the need for post-selection in the preparation of $\ket{\tilde{\epsilon}_0}$ with the fixed-point amplitude amplification algorithm \cite{FixedPointSearch2014}.
Deterministic state preparation is essential to employ advanced expectation value estimation techniques such as RAE \cite{RAE-first,RAE-VQE,RAEexperiment}.
When applied to our setting, it allows us to deterministically prepare a state 
\begin{align}\label{eq:fixed_point_aa}
    \ket{\tilde{\tilde{\epsilon}}_{0}} = \sqrt{1-r^2}\ket{0^{l'+1}}\ket{\tilde{\epsilon}_{0}} + r\ket{0\tilde{\epsilon}_0^\perp},
\end{align}
where $\ket{0\tilde{\epsilon}_0^\perp}$ is a state orthogonal to $\ket{0^{l'+1}}\ket{\tilde{\epsilon}_{0}}$, for $r>0$ with approximately
\begin{align}\label{eq:fixedpointaacost}
    \frac{\log_e(2/r)}{\sqrt{p}}
\end{align}
applications of $\mathcal{P}^{\mathrm{filter}}_{\varepsilon,\kappa,x_{\mathrm{th}}}$ and $(\mathcal{P}^{\mathrm{filter}}_{\varepsilon,\kappa,x_{\mathrm{th}}})^\dagger$ \cite{FixedPointSearch2014}.
Note that expectation value of $V$ and $V\Pi(H-\epsilon_0)^{-1}\Pi V$ with respect to $\ket{\tilde{\tilde{\epsilon}}_0}$ deviates from $\braket{\tilde{\epsilon}_0|V|\tilde{\epsilon}_0}$ by $2r\mathrm{Re}[\braket{0\tilde{\epsilon}_0|V|0\tilde{\epsilon}_0^\perp}]+\mathcal{O}(r^2)$ and $2r\mathrm{Re}[\braket{0\tilde{\epsilon}_0|V\Pi(H-\epsilon_0)^{-1}\Pi V|0\tilde{\epsilon}_0^\perp}]+\mathcal{O}(r^2)$.
It can roughly be bounded from above by $2r\|V\|$ and $2r\|V\|^2/\Delta$.
We need to take sufficiently small $r$ to make this negligible.

Finally, let us discuss the related previous works about ground state preparation.
Ge, Tura, and Cirac \cite{Ge2017filtering} also consider how to realize filtering operations but with a so-called linear combination of unitaries approach which originates in \cite{Childs2017QLSP}.
The approach requires $\log_2(d)$ ancilla qubits to implement degree-$d$ polynomial.
Refs. \cite{Lin2020nearoptimalground, FTQC-derivative} use QSP to implement a reflection operator $I - 2\ket{\epsilon_0}\bra{\epsilon_0}$, and uses the fixed-point amplitude amplification.
They do not perform a detailed error analysis of the protocol as we do in this paper; we expect that it would be more involved and that such analysis would yield a comparable performance to the method presented in this section.
All of the above works give the cost in $\mathcal{O}$-notation and do not give detailed cost estimates as we do in this work.

\subsection{First-order perturbation energy}\label{sec:first-order}

The first-order energy correction $\epsilon_0^{(1)}$ can be obtained by naive measurement of the operator $V$, that is, we perform the VQE-like measurement where we estimate each of Pauli operator $\sigma_\ell$ appearing in Eq. \eqref{eq:V-def} by the RAE.

The technique demands us to deterministically prepare $\ket{\epsilon_0}$.
To this end, we employ the fixed-point amplitude amplification algorithm \cite{FixedPointSearch2014}.
Let the unitary that prepares $\ket{\tilde{\tilde{\epsilon}}_{0}}$ in Eq. \eqref{eq:fixed_point_aa} be $W_{\mathrm{prep}}$.
Then, the RAE can empirically estimate $\braket{\tilde{\tilde{\epsilon}}_{0}|\sigma_\ell|\tilde{\tilde{\epsilon}}_{0}}$ with a mean squared error $\delta^2_{\sigma_\ell}$ using 
\begin{align}
    M_\ell = \frac{5\sqrt{2}}{2}\frac{e^2}{e-1}\frac{1}{\delta_{\sigma_\ell}}
\end{align} 
calls of $W_{\mathrm{prep}}$ in total \cite[Eq. (8)]{RAE-VQE} (converted here to be a noise-free case).
Employing optimal distribution of $M_\ell$ to minimize the overall error, we can obtain an estimate of $\braket{\tilde{\tilde{\epsilon}}_{0}|V|\tilde{\tilde{\epsilon}}_{0}}$ with a mean squared error $\delta_1^2$ using
\begin{align}\label{eq:total-cost-RAE}
    M^{(1)} = \frac{5\sqrt{2}}{2}\frac{e^2}{e-1}\frac{1}{\delta_1}\left(\sum_\ell v_\ell^{2/3}\right)^{3/2}
\end{align}
calls of $W_{\mathrm{prep}}$ in total.
For completeness, we derive Eq.~\eqref{eq:total-cost-RAE} in Appendix \ref{appsec:optimal-dist-RAE} following Ref. \cite{RAE-VQE} where they have derived the total cost for noisy quantum computers.

Here, we combine the above discussion to state a formal result in the $\mathcal{O}$-notation as follows:
\begin{theorem}\label{thm:first-order}
    Let $H_{\mathrm{total}}= H+V$ be the target Hamiltonian where $H = \sum_{\ell=1}^{L_H} h_\ell \sigma_\ell$ represents the unperturbed term and $V = \sum_{\ell=1}^{L_V} v_\ell \sigma_\ell$ represents the perturbation. Let $\ket{\epsilon_0}$ be the non-degenerate ground state of $H$. Assume that we have an estimate $\hat{\epsilon}_0$ of $\epsilon_0$, the ground state energy of $H$, such that $|\hat{\epsilon}_0-\epsilon_0|<\delta_0<\Delta$. Moreover, assume that we can preprare a state $\ket{\psi}$ such that $|\braket{\epsilon_0|\psi}|^2=p$. Then, we can estimate the first-order perturbation energy $\braket{\epsilon_0|V|\epsilon_0}$ within an additive error of $\delta_1$ by using
    \begin{align}
        \mathcal{O}\left(\frac{\|\bm{h}'\|_1\|\bm{v}\|_{2/3}}{\Delta'\delta_1\sqrt{p}}\log\left(\frac{\|V\|}{\delta_1}\right)\log\left(\sqrt{\frac{p}{1-p}}\frac{\|V\|}{\delta_1}\right)\right)
    \end{align}
    calls of a block-encoding of $H'$, where $\Delta'=\Delta-\delta_0$ and $H'=H-\hat{\epsilon}_0$.
\end{theorem}
\textit{Proof.} This is obtained by multiplying Eq.~\eqref{eq:total-cost-RAE}, Eq.~\eqref{eq:fixedpointaacost} with $r=\mathcal{O}(\delta_1/\|V\|)$ and Eq.~\eqref{eq:filtering-degree-approx-upper-bound} with setting $\kappa$ and $x_{\mathrm{th}}'$ as in Eqs.~\eqref{eq:kappa}-\eqref{eq:x_th} and $\varepsilon=\mathcal{O}(\sqrt{p/(1-p)}\delta_1/\|V\|)$ (c.f. Eq.~\eqref{eq:delta_prep_first_order}). \hfill$\square$

We will estimate the concrete number of calls to the block-encoding of $H'$ in Sec.~\ref{sec:resource-analysis}.

\subsection{Second-order perturbation energy}\label{sec:second-order}
\begin{figure}
    \centering
    \includegraphics[width=0.85\linewidth]{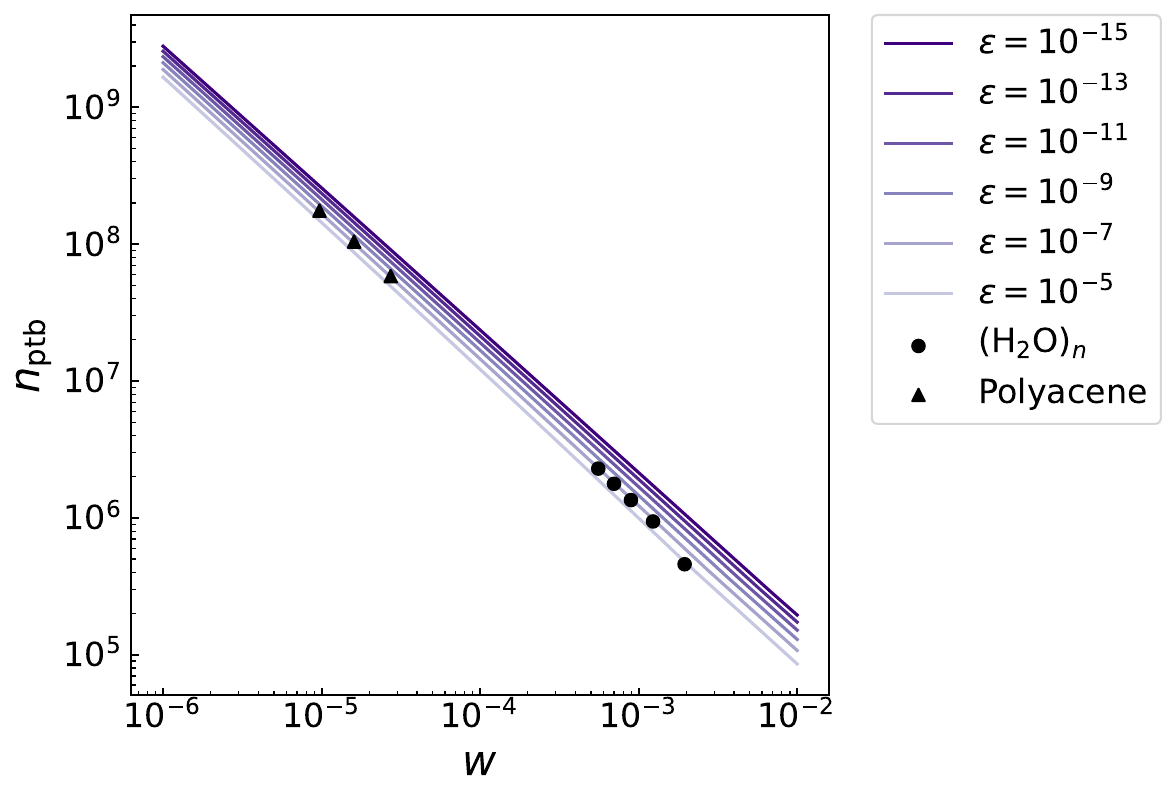}
    \caption{\label{fig:ptb} Values of $n_{\mathrm{ptb}}$ calculated by Eq.~\eqref{eq:n-ptb} with $w_0=10^{-6}$, which is a typical value for the molecules studied in Sec.~\ref{sec:resource-analysis}, and with different error parameters $\varepsilon$ as a function of $w$. Points corresponds to the values for specific molecules presented in Table \ref{tab:second_order_perturbation}.}
\end{figure}

We approximate $\Pi(H-\epsilon_0)^{-1}\Pi$ by the method presented in \cite{GrandUnification}.
Ref. \cite{GrandUnification} gives a construction of a QSP-implementable polynomial to realize matrix inversion.
However, as we show in Appendix \ref{appsec:perturbation}, it can also be used for approximating $\Pi(H-\epsilon_0)^{-1}\Pi$.
More concretely, there exists a QSP-implementable polynomial $P^{\mathrm{ptb}}_{\varepsilon, w, w_0}(x)$ that satisfies the following conditions:
\begin{align}
    &\left|P^{\mathrm{ptb}}_{\varepsilon, w, w_0}(x)-\frac{w}{2}\frac{1}{x}\right|<\frac{w}{2}\varepsilon~~(w<|x|<1) \label{eq:Pptb_inv}\\
    &\left|P^{\mathrm{ptb}}_{\varepsilon, w, w_0}(x)\right|<\frac{w}{2}\varepsilon~~(|x|<w_0) \label{eq:Pptb_zero}
\end{align}
Such $P^{\mathrm{ptb}}_{\varepsilon, w, w_0}(x)$ can be constructed by appropriately tuning the error parameter of a matrix inversion polynomial presented in \cite{GrandUnification}, as shown in Appendix \ref{appsec:perturbation}.
Its degree is
\begin{align}\label{eq:n-ptb}
    n_{\mathrm{ptb}}(\varepsilon, w, w_0) &= 2D\left(\frac{\varepsilon}{4}, \frac{w}{2}\right) + n_{\mathrm{sign}}\left(\varepsilon'', \frac{w}{4}, \frac{3w}{4}\right),
\end{align}
where
\begin{align}
    &D(\varepsilon, w) =\left\lceil\sqrt{b(\varepsilon, w) \log (4 b(\varepsilon, w) / \varepsilon)}\right\rceil, \\
        \begin{split}
            &n_{\mathrm{sign}}(\varepsilon, \kappa, c)
            \\
            &= \frac{32 (1+|c|)}{\sqrt{\pi} \varepsilon} \frac{1}{\kappa}\sqrt{2\log_e \left(\frac{8}{\pi\varepsilon^2}\right)} \exp\left(- \frac{1}{2}W\left(\frac{2048}{\pi \varepsilon^{2} e^{2}}\right)\right)
        \end{split}\\
    &\varepsilon'' = \min\left\{\frac{2\varepsilon w}{5},~\frac{1}{4wD(\frac{\varepsilon}{4},\frac{w}{2})},~\frac{\varepsilon}{2w_0 [D(\frac{\varepsilon}{4},\frac{w}{2})+1]^2}\right\},
\end{align}
and
\begin{align}
    b(\varepsilon, w) = \left\lceil\left(\frac{1}{w}\right)^{2} \log \left(\frac{1}{w\varepsilon}\right)\right\rceil. \label{eq:b}
\end{align}
We plot the values of $n_{\mathrm{ptb}}(\varepsilon, w, w_0)$ as Fig.~\ref{fig:ptb} with various parameter settings.

Let $\mathcal{P}^{\mathrm{ptb}}_{\varepsilon, w, w_0}$ be a unitary that implements $P^{\mathrm{ptb}}_{\varepsilon, w, w_0}(H'/\|\bm{h}'\|_1)$, that is, for a general state $\ket{0^{l'+1}}\ket{\psi}$, it acts as,
\begin{align}
    \begin{split}
        &\mathcal{P}^{\mathrm{ptb}}_{\varepsilon, w, w_0}\ket{0^{l'+1}}\ket{\psi}\\
        &= \ket{0^{l'+1}}\left[P^{\mathrm{ptb}}_{\varepsilon, w, w_0}\left(\frac{H'}{\|\bm{h}'\|_1}\right)\ket{\psi}\right] + \ket{g}        
    \end{split}\\
    &\approx \ket{0^{l'+1}}\left[\frac{w}{2} \|\bm{h}'\|_1\Pi\left(H-\epsilon_0\right)^{-1}\Pi\ket{\psi}\right] + \ket{g} 
\end{align}
$\epsilon_0^{(2)}$ can be approximated by the expectation value of the operator $\mathcal{P}^{\mathrm{ptb}}_{\varepsilon, w, w_0}$ as,
\begin{align}\label{eq:second-order-perturbation-energy}
    \epsilon_0^{(2)} \approx \tilde{\epsilon}_0^{(2)} := -\frac{2}{w}\frac{1}{\|\bm{h}'\|_1}\braket{\mathcal{P}^{\mathrm{ptb}}_{\varepsilon, w, w_0}}
\end{align}
where the expectation is taken with respect to $\ket{0^{l'+1}}(V\ket{\epsilon_0})$.
In Appendix \ref{appsec:second-order-error}, we show that, by taking
\begin{align}
    w&=\frac{\Delta-\delta_0}{\|\bm{h}'\|_1}, \label{eq:w-concrete} \\
    w_0&=\frac{\delta_0}{\|\bm{h}'\|_1}, \label{eq:w0-concrete}
\end{align}
it is guaranteed that
\begin{align}
    \left|\epsilon_{0}^{(2)} - \tilde{\epsilon}_{0}^{(2)}\right| \leq \frac{\|V\ket{\epsilon_0}\|^2 }{\|\bm{h}'\|_1}\varepsilon + \frac{\delta_0}{(\Delta-\delta_0)}\epsilon_{0}^{(2)}. 
    \label{eq:second-order-error}
\end{align}
Let us define
\begin{align}
    \delta_2' &= \frac{\|V\ket{\epsilon_0}\|^2 }{\|\bm{h}'\|_1}\varepsilon \\
    \delta_2'' &= \frac{\delta_0}{(\Delta-\delta_0)}\epsilon_{0}^{(2)}
\end{align}
If we wish to obtain the overall perturbation energy within an error of $\delta$, we have to make Eq. \eqref{eq:second-order-error} sufficiently smaller than $\delta$ by taking small $\varepsilon$ and $\delta_0$.
As for $\varepsilon$, we can for example take $\varepsilon=(\|\bm{h}'\|_1/\|V\|^2)(\delta/10)$ without increasing the implementation cost so much to guarantee that $\delta_2'$ is negligible with respect to $\delta$.
Note that $1/\varepsilon$ only contributes logarithmically to $n_{\mathrm{ptb}}$.
As for $\delta_0$, although its effect to $n_{\mathrm{ptb}}$ is logarithmic, the phase estimation of $H$ takes $\frac{\sqrt{2}\pi \|\bm{h}\|_1}{2\delta_0}$ calls of $U_H$.
We therefore wish to take $\delta_0$ as large as possible while maintaining $\frac{\delta_0}{(\Delta-\delta_0)}\epsilon_{0}^{(2)}\ll \delta$.

Now, we discuss how to obtain an estimate of $\braket{\mathcal{P}^{\mathrm{ptb}}_{\varepsilon, w, w_0}}$ and therefore $\tilde{\epsilon}_0^{(2)}$.
One strategy to measure $\braket{\mathcal{P}^{\mathrm{ptb}}_{\varepsilon, w, w_0}}$ is to generate $V\ket{\epsilon_0}$ via the block-encoding of $V$.
This, however, would prevent us from analyzing the contribution of each term to the perturbative energy. 
We take a more naive strategy to avoid this issue.
Since $V$ can be expressed as Eq. \eqref{eq:V-def}, $\braket{\mathcal{P}^{\mathrm{ptb}}_{\varepsilon, w, w_0}}$ can be rewritten as,
\begin{align}
    \braket{\mathcal{P}^{\mathrm{ptb}}_{\varepsilon, w, w_0}} = \sum_{\ell, \ell'} v_\ell v_{\ell'} \braket{\mathcal{P}^{\mathrm{ptb}}_{\varepsilon, w, w_0}}_{\ell'\ell}.
\end{align}
where
\begin{align}
    \braket{\mathcal{P}^{\mathrm{ptb}}_{\varepsilon, w, w_0}}_{\ell'\ell} = \bra{0^{l'+1}}\bra{\epsilon_0} \sigma_{\ell'}\mathcal{P}^{\mathrm{ptb}}_{\varepsilon, w, w_0}\sigma_{\ell}\ket{0^{l'+1}}\ket{\epsilon_0}
\end{align}
We can estimate $\braket{\mathcal{P}^{\mathrm{ptb}}_{\varepsilon, w, w_0}}_{\ell'\ell}$ using Hadamard test involving a controlled-$(\sigma_{\ell'}\mathcal{P}^{\mathrm{ptb}}_{\varepsilon, w, w_0}\sigma_{\ell})$ gate.
Note that controlled-$\mathcal{P}^{\mathrm{ptb}}_{\varepsilon, w, w_0}$ does not increase the T cost by much; it only adds a single control qubit to the controlled-$U_{H'}$ gates already used to realize $\mathcal{P}^{\mathrm{ptb}}_{\varepsilon, w, w_0}$.
We first deterministically generate $\ket{0^{l'+1}}\ket{\tilde{\epsilon}_0}$ using filtering described in Sec. \ref{sec:ground-state-preparation}.
Then, adding an ancilla qubit, we prepare a state
\begin{align}\label{eq:second-order-pert-hadamard-test}
        \frac{1}{\sqrt{2}}\ket{0}\left(\ket{0^{l'+1}}\ket{\tilde{\epsilon}_0}\right) 
    +\frac{1}{\sqrt{2}}\ket{1}\left(\sigma_{\ell'}\mathcal{P}^{\mathrm{ptb}}_{\varepsilon, w, w_0}\sigma_{\ell}\ket{0^{l'+1}}\ket{\tilde{\epsilon}_0}\right).
\end{align} 
Pauli-$X$ expectation value of the additional ancilla qubit of this state is $\mathrm{Re}\bra{0^{l'+1}}\bra{\epsilon_0} \sigma_{\ell'}\mathcal{P}^{\mathrm{ptb}}_{\varepsilon, w, w_0}\sigma_{\ell}\ket{0^{l'+1}}\ket{\epsilon_0}=\braket{\mathcal{P}^{\mathrm{ptb}}_{\varepsilon, w, w_0}}_{\ell'\ell}$, which is the quantity we wish to estimate.
This allows us to employ the RAE to estimate the expectation values $\braket{\mathcal{P}^{\mathrm{ptb}}_{\varepsilon, w, w_0}}_{\ell'\ell}$.
Denoting the unitary to prepare the state in Eq. \eqref{eq:second-order-pert-hadamard-test} by $W_2$, the optimal number of calls of $W_2$ to estimate $\braket{\mathcal{P}^{\mathrm{ptb}}_{\varepsilon, w, w_0}}$ with a mean squared error $\delta^2_{\mathcal{P}}$ is,
\begin{align}
    M^{(2)}_{\mathcal{P}} &= \frac{5\sqrt{2}}{2}\frac{e^2}{e-1}\frac{1}{\delta_{\mathcal{P}}}\left(\sum_{\ell\ell'} (v_\ell v_{\ell'})^{2/3}\right)^{3/2} \\
    &=\frac{5\sqrt{2}}{2}\frac{e^2}{e-1}\frac{1}{\delta_{\mathcal{P}}}\left(\sum_{\ell} v_\ell^{2/3}\right)^{3}
\end{align}
which follows from exactly the same discussion provided in Appendix \ref{appsec:optimal-dist-RAE}.
Since $\tilde{\epsilon}_0^{(2)}$ is obtained by multiplying $2/(w\|\bm{h}'\|_1)$ to the estimated value of $\braket{\mathcal{P}^{\mathrm{ptb}}_{\varepsilon, w, w_0}}$, we need to take $\delta_{\mathcal{P}} = \frac{w\|\bm{h}'\|_1}{2}\delta_2$ to obtain $\tilde{\epsilon}_0^{(2)}$ with a mean squared error ${\delta_2}^2$.
Therefore, we conclude that,
\begin{align}\label{eq:total-cost-RAE-second-order}
    M^{(2)} &=\frac{5\sqrt{2}e^2}{e-1}\frac{1}{w\|\bm{h}'\|_1}\frac{1}{\delta_2}\left(\sum_{\ell} v_\ell^{2/3}\right)^{3}
\end{align}
calls of $W_2$ are needed to estimate $\tilde{\epsilon}_0^{(2)}$.

Combining the above discussion, we obtain the following result:
\begin{theorem}
    Let $H_{\mathrm{total}}$, $H$, $V$, $h_\ell$, $v_\ell$, $\ket{\epsilon_0}$ $\hat{\epsilon}_0$, $\delta_0$, $\ket{\psi}$, $\Delta'$, $H'$ and $p$ be defined as in Theorem \ref{thm:first-order}. Additionally, let $\Pi=I-\ket{\epsilon_0}\bra{\epsilon_0}$. Then, we can estimate the second-order perturbation energy $-\bra{\epsilon_0}V \Pi (H-\epsilon_0)^{-1}\Pi V \ket{\epsilon_0}$ within an additive error of $\delta_2$ by using
    \begin{align}
        \mathcal{O}\left(\frac{\|\bm{h}\|_1\|\bm{v}\|_{2/3}^2}{{\Delta'}^2\delta_2\sqrt{p}}\log\left(\frac{\|V\|^2}{\Delta\delta_2}\right)\log\left(\sqrt{\frac{p}{1-p}}\frac{\|V\|^2}{\Delta\delta_2}\right)\right)
    \end{align}
    calls of a block-encoding of $H'$.
\end{theorem}
\textit{Proof.} This is obtained by multiplying the following equations.
\begin{itemize}
    \item Eq.~\eqref{eq:total-cost-RAE-second-order},
    \item Eq.~\eqref{eq:fixedpointaacost} with $r=\mathcal{O}(\delta_2\Delta/\|V\|^2)$,
    \item Eq.~\eqref{eq:filtering-degree-approx-upper-bound} with setting $\kappa$ and $x_{\mathrm{th}}'$ as in Eqs.~\eqref{eq:kappa}-\eqref{eq:x_th} and $\varepsilon=\mathcal{O}(\sqrt{p/(1-p)}\delta_2\Delta/\|V\|^2)$ (c.f. Eq.~\eqref{eq:delta_prep_second_order}) and
    \item Eq.~\eqref{eq:n-ptb} with $\varepsilon=\mathcal{O}(\delta_2\|\bm{h}'\|_1/\|V\|^2)$ (c.f. Eq.\eqref{eq:second-order-error}) and $w$ as defined in Eq.~\eqref{eq:w-concrete}. \hfill$\square$
\end{itemize}

We will calculate the concrete values in the next section.

\section{Resource analysis with realistic parameters}\label{sec:resource-analysis}

Here, we assess the overall cost for obtaining the ground state energy $E_0$ within the accuracy of $\delta$ by the perturbative method presented in the previous section.
We analyze a rough T-cost of our approach with an assumption that the second-order perturbation is accurate enough to produce $E_0$.
Let the T-cost of implementing $U_H$ be $\mathcal{T}(H)$.
After formulating the overall cost of the proposed approach, we plug in some realistic parameters to the formula and discuss its feasibility.

\subsection{Formulating the overall cost}\label{sec:overall-cost}
Let us first discuss how to distribute the overall error $\delta$ to $\delta_0$, $\delta_1$, and $\delta_2$.
Importantly, $\delta_2''$ depends on $\delta_0$.
Assuming $\delta_0\ll\Delta$, $\delta_2''$ can be approximated as $\frac{\epsilon_0^{(2)}}{\Delta}\delta_0$.
$\delta_2''$ is negligibly small compared to $\delta_0$ and thus to $\delta$ if $\epsilon_0^{(2)}\ll\Delta$, which is usually satisfied when, e.g., $V$ represents intermolecular interactions.
In this situation, we can naively set $\delta_0=\delta_1=\delta_2=\delta/3$ to guarantee that the overall accuracy is $\delta$.
Conversely, if $\epsilon_0^{(2)}\gg\Delta$, we must take $\delta_0 \ll \frac{\Delta}{\epsilon_0^{(2)}}\delta$.
In this situation, we can set $\delta_1=\delta_2 = \delta_2'' = \delta/3$ since $\delta_0 = \frac{\Delta}{\epsilon_0^{(2)}}\delta_2''$ becomes negligibly small compared to $\delta$.
Note that we may reduce the cost further by optimizing this distribution, which is left as possible future work.

The contribution to the T-cost needed for each step of the presented perturbation method can be summarized as follows:
    \begin{description}
        \item[Estimation of $\epsilon_0$]\mbox{}\\
        Phase estimation of $H$ to obtain $\epsilon_0$ within an accuracy of $\delta_0$ takes
        \begin{align}
            \frac{1}{p}\frac{\sqrt{2}\pi \|\bm{h}\|}{2\delta_0}\mathcal{T}(H),
        \end{align}
        since we can obtain the ground state energy with probability $p$ if we use an initial state \eqref{eq:initial-state}. Note that if we expect $\epsilon_0^{(2)}\gg\Delta$, we must take $\delta_0 \ll \frac{\Delta}{\epsilon_0^{(2)}}\delta$.
        
        \item[Estimation of $\epsilon_0^{(1)}$]\mbox{}\\
        It takes $M^{(1)}$ (Eq. \eqref{eq:total-cost-RAE}) calls of $W_{\mathrm{prep}}$ which uses $\frac{\log_e(2/r)}{\sqrt{p}}$ calls of $\mathcal{P}^{\mathrm{filter}}_{\varepsilon,\kappa,x_{\mathrm{th}}}$ and ${\mathcal{P}^{\mathrm{filter}}_{\varepsilon,\kappa,x_{\mathrm{th}}}}^\dagger$. $\mathcal{P}^{\mathrm{filter}}_{\varepsilon,\kappa,x_{\mathrm{th}}}$ takes $n_{\mathrm{filter}}(\varepsilon, \kappa, x_{\mathrm{th}})\mathcal{T}(H)$ cost for its implementation.
        Then, the overall cost is
        \begin{align}\label{eq:overall-cost-first-order}
            2M^{(1)}\frac{\log_e(2/r)}{\sqrt{p}}n_{\mathrm{filter}}(\varepsilon, \kappa, x_{\mathrm{th}})\mathcal{T}(H)
        \end{align}
        The parameters in Eq. \eqref{eq:overall-cost-first-order} are determined as follows:
        \begin{itemize}
            \item $r$ should be taken to make $2r\|V\|$ negligible compared to $\delta_1$. Noting that $\|V\|\leq\|\bm{v}\|_1$ holds, we take $r=\delta_1/(20\|\bm{v}\|_1)$ which guarantees $2r\|V\|\leq\delta_1/10$. 
            \item We should take $\varepsilon$ to make $\delta_{\mathrm{prep}}$ negligible to $\delta_1$. From Eq. \eqref{eq:delta_prep_first_order}, we choose $\varepsilon=\delta_1/(20\|\bm{v}\|_1)\sqrt{p/(1-p)}$ which leads to $\delta_{\mathrm{prep}}\lesssim\delta_1/10$.
            \item $\kappa$ and $x_{\mathrm{th}}$ are determined by Eqs. \eqref{eq:x_th} and \eqref{eq:kappa}.
        \end{itemize}

        \item[Estimation of $\epsilon_0^{(2)}$]\mbox{}\\
        It takes $M^{(2)}$ (Eq. \eqref{eq:total-cost-RAE-second-order}) calls of $W_2$ which makes use of single calls of $W_{\mathrm{prep}}$ and $\mathcal{P}^{\mathrm{ptb}}_{\varepsilon, w, w_0}$. 
        The cost of $W_{\mathrm{prep}}$ is $2\frac{\log_e(2/r)}{\sqrt{p}}n_{\mathrm{filter}}(\varepsilon_{\mathrm{filter}}, \kappa, x_{\mathrm{th}})\mathcal{T}(H)$. That of $\mathcal{P}^{\mathrm{ptb}}_{\varepsilon_{\mathrm{ptb}}, w, w_0}$ is $n_{\mathrm{ptb}}(\varepsilon_{\mathrm{ptb}}, w, w_0)\mathcal{T}(H)$. $n_{\mathrm{ptb}}(\varepsilon_{\mathrm{ptb}}, w, w_0)$ is defined by Eqs. \eqref{eq:n-ptb}-\eqref{eq:b}.
        Therefore, the overall cost is,
        \begin{align}
            M^{(2)}&\bigg(2\frac{\log_e(2/r)}{\sqrt{p}}n_{\mathrm{filter}}(\varepsilon_{\mathrm{filter}}, \kappa, x_{\mathrm{th}}) \nonumber \\
            &\quad +n_{\mathrm{ptb}}(\varepsilon_{\mathrm{ptb}}, w, w_0)\bigg)\mathcal{T}(H)
            \end{align}            
        The parameters are determined as follows:
        \begin{itemize}
            \item $r$ should be taken to make $2r\|V\|^2/\Delta$ negligible compared to $\delta_2$. We take $r=\delta_2\Delta/(20\|\bm{v}\|_1^2)$ which guarantees $2r\|V\|\leq\delta_2/10$. 
            \item We take $\varepsilon_{\mathrm{filter}}=\delta_2\Delta/(20\|\bm{v}\|_1^2)\sqrt{p/(1-p)}$ which leads to $\delta_{\mathrm{prep}}\lesssim\delta_2/10$ (see Eq. \ref{eq:delta_prep_second_order}).
            \item $\kappa$ and $x_{\mathrm{th}}$ are determined by Eqs. \eqref{eq:x_th} and \eqref{eq:kappa}.
            \item $\varepsilon_{\mathrm{ptb}}$ should be taken to make $\delta_2' = \frac{\|V\ket{\epsilon_0}\|^2 }{\|\bm{h}'\|_1}\varepsilon_{\mathrm{ptb}}\leq\frac{\|V\|^2 }{\|\bm{h}'\|_1}\varepsilon_{\mathrm{ptb}}$ sufficiently smaller than $\delta_2$ (see Eq. \eqref{eq:second-order-error}). We therefore take $\varepsilon_{\mathrm{ptb}}=\frac{\|\bm{h}'\|_1}{\|\bm{v}\|^2_1}\frac{\delta_2}{10}$ to ensure $\delta_2' \leq \delta_2/10$.
            \item $w$ and $w_0$ are determined by Eq. \eqref{eq:w-concrete} and \eqref{eq:w0-concrete}.
        \end{itemize}
    \end{description}        

\subsection{Application to molecular systems}

\begin{table*}
    \caption{\label{tab:properties}Parameters of example systems. For $(\mathrm{H_2O})_n$, $\Delta$ is computed as the exact energy gap of a single water molecule isolated from the other ones. Energy units are in Hartree.}
    \centering
        \begin{tabular}{lllllll}
            \hline\hline
            System & $\|\bm{h}'\|_1$ & $\|\bm{v}\|_1$ & $\left(\sum_{\ell}v_\ell^{2/3}\right)^{3/2}$ & $\Delta$ & $p$ & $L_V/L_H$ \\
            $(\mathrm{H_2O})_2$ & 204 & 14.3 & 783 & 0.40 & 0.97 & 7.18 \\
            $(\mathrm{H_2O})_3$ & 323 & 197 & 1.1$\times 10^4$ & 0.40 & 0.97 & 27.6 \\
            $(\mathrm{H_2O})_4$ & 445 & 313 & 2.1$\times 10^4$ & 0.40 & 0.97 & 66.3 \\
            $(\mathrm{H_2O})_5$ & 570 & 438 & 3.6$\times 10^4$ & 0.40 & 0.97 & 129 \\
            $(\mathrm{H_2O})_6$ & 715 & 610 & 7.0$\times 10^4$ & 0.40 & 0.97 & 224 \\
            
            Tetracene & 2.0$\times 10^3$ & 9.5 $\times 10^3$ & 3.2$\times 10^7$ & 0.055 & 0.7* & 323 \\
            Pentacene & 2.7$\times 10^3$ & 1.5 $\times 10^4$ & 7.4$\times 10^7$ & 0.043 & 0.7* & 348 \\
            Hexacene & 3.4$\times 10^3$ & 2.3 $\times 10^4$ & 1.5$\times 10^8$ & 0.033 & 0.7* & 385\\
            \hline\hline 
        \end{tabular}
\end{table*}

\begin{table*}
    \caption{\label{tab:first_order_perturbation} Total cost for estimating $\epsilon^{(1)}_0$ and intermediate parameters used during its calculation.}
    \centering
        \begin{tabular}{llllllllll}
            \hline\hline
            System & $r$ & $\varepsilon_{\mathrm{filter}}$ & $\kappa$ & $n_{\mathrm{filter}}$ & $M^{(1)}$ & Total cost \\
            $(\mathrm{H_2O})_2$ & 5.2$\times 10^{-7}$ & 3.2$\times 10^{-6}$ & 1.9$\times 10^{-3}$ & 7.4$\times 10^{4}$ & $1.2\times 10^{4}$ & 5.4$\times 10^{10}$$\mathcal{T}(H)$\\

            $(\mathrm{H_2O})_3$ & 2.5$\times 10^{-8}$ & 1.5$\times 10^{-7}$ & 1.2$\times 10^{-3}$ & 1.4$\times 10^{5}$ & $1.6\times 10^{5}$ & 2.6$\times 10^{12}$$\mathcal{T}(H)$\\
            
            $(\mathrm{H_2O})_4$ & 1.2$\times 10^{-8}$ & 7.3$\times 10^{-8}$ & 8.9$\times 10^{-4}$ & 2.1$\times 10^{5}$ & $3.2\times 10^{5}$ & 1.0$\times 10^{13}$$\mathcal{T}(H)$\\
            
            $(\mathrm{H_2O})_5$ & 6.8$\times 10^{-9}$ & 4.2$\times 10^{-8}$ & 7.0$\times 10^{-4}$ & 2.7$\times 10^{5}$ & $5.6\times 10^{5}$ & 3.0$\times 10^{13}$$\mathcal{T}(H)$\\
            
            $(\mathrm{H_2O})_6$ & 4.1$\times 10^{-9}$ & 2.5$\times 10^{-8}$ & 5.6$\times 10^{-4}$ & 3.5$\times 10^{5}$ & $1.1\times 10^{6}$ & 9.1$\times 10^{13}$$\mathcal{T}(H)$\\
            \hline\hline



        \end{tabular}
\end{table*}

\begin{table*}
    \caption{\label{tab:second_order_perturbation} Total cost for estimating $\epsilon^{(2)}_0$ and intermediate parameters used during its calculation.}
    \centering
        \begin{tabular}{llllllllll}
            \hline\hline
            System & $r$ & $\varepsilon_{\mathrm{filter}}$ & $\varepsilon_{\mathrm{ptb}}$ & $n_{\mathrm{filter}}$ & $n_{\mathrm{ptb}}$ & $M^{(2)}$ & Total cost \\
            $(\mathrm{H_2O})_2$ & 1.4$\times 10^{-8}$ & 8.8$\times 10^{-8}$ & 3.0$\times 10^{-5}$ & 9.3$\times 10^{4}$ & $4.6\times 10^{5}$ & 1.6$\times 10^{11}$ & 1.2$\times 10^{18}\mathcal{T}(H)$\\

            $(\mathrm{H_2O})_3$ & 5.1$\times 10^{-11}$ & 3.0$\times 10^{-10}$ & 1.8$\times 10^{-7}$ & 2.0$\times 10^{5}$ & $9.4\times 10^{5}$ & 2.9$\times 10^{13}$ & 8.8$\times 10^{20}\mathcal{T}(H)$\\

            $(\mathrm{H_2O})_4$ & 1.5$\times 10^{-11}$ & 9.2$\times 10^{-11}$ & 9.8$\times 10^{-8}$ & 2.8$\times 10^{5}$ & $1.3\times 10^{6}$ & 1.1$\times 10^{14}$ & 6.8$\times 10^{21}\mathcal{T}(H)$\\
            
            $(\mathrm{H_2O})_5$ & 6.2$\times 10^{-12}$ & 3.7$\times 10^{-11}$ & 6.5$\times 10^{-8}$ & 3.8$\times 10^{5}$ & $1.8\times 10^{6}$ & 3.4$\times 10^{14}$ & 3.6$\times 10^{22}\mathcal{T}(H)$\\
            
            $(\mathrm{H_2O})_6$ & 2.7$\times 10^{-12}$ & 1.6$\times 10^{-11}$ & 4.2$\times 10^{-8}$ & 4.9$\times 10^{5}$ & $2.3\times 10^{6}$ & 1.3$\times 10^{15}$ & 2.1$\times 10^{23}\mathcal{T}(H)$\\

            Tetracene & 9.2$\times 10^{-15}$ & 1.4$\times 10^{-14}$ & 6.7$\times 10^{-10}$ & 1.3$\times 10^{7}$ & $5.8\times 10^{7}$ & 1.9$\times 10^{21}$ & 2.0$\times 10^{30}\mathcal{T}(H)$\\

            Pentacene & 2.8$\times 10^{-15}$ & 4.2$\times 10^{-15}$ & 3.4$\times 10^{-10}$ & 2.3$\times 10^{7}$ & $1.0\times 10^{8}$ & 1.3$\times 10^{22}$ & 2.6$\times 10^{31}\mathcal{T}(H)$\\
            
            Hexacene & 9.6$\times 10^{-16}$ & 1.5$\times 10^{-15}$ & 2.0$\times 10^{-10}$ & 3.9$\times 10^{7}$ & $1.8\times 10^{8}$ & 6.8$\times 10^{22}$ & 2.3$\times 10^{32}\mathcal{T}(H)$\\
            \hline\hline
        \end{tabular}
\end{table*}

We first perform resource estimation for clusters of water molecule $(\mathrm{H_2O})_m$ for $m=2,3,4,5,6$.
Geometries of systems are taken from Ref. \cite{WALES199865}.
We use the STO-3G minimal basis set to represent the Hamiltonians, which are computed with PySCF \cite{pyscf2018,pyscf2020} and OpenFermion \cite{openfermion}.
Hamiltonians are expressed L\"owdin localized orbital \cite{lowdin1950non}, which allows us to separate them into intra-molecule and inter-molecule interactions.
Here, we estimate the cost to implement the perturbation theory presented in the previous section treating the inter-molecule interactions as the perturbative term $V$.
Since we expect $\epsilon_0^{(2)}\ll\Delta$ in this case, we set $\delta_0=\delta_1=\delta_2=\delta_{\mathrm{chem}}/3$ and $\delta_{\mathrm{chem}}=10^{-3}$ Hartree.

For a cluster of $m$ molecules like the ones considered here, a special treatment for generating the unperturbed ground state $\ket{\epsilon_0}$ can be made. 
The unperturbed Hamiltonian $H$ can be written in the form of,
\begin{align}
    H &= \sum_{i=1}^m H_i,
\end{align}
where $H_i$ only acts on qubits corresponding to localized orbitals in the $i$-th molecule.
Note that the above form is always possible with appropriate indexing of orbitals and Jordan-Wigner transformation \cite{Jordan1928}.
Then, the ground state $\ket{\epsilon_0}$ of $H$ is a tensor product of ground states $\ket{\epsilon_{0,i}}$ of $H_i$.
Let us now assume that we can efficiently generate $\ket{\psi_i}$ such that $\sqrt{p_i}\ket{\epsilon_{0,i}}+\sqrt{1-p_i}\ket{\epsilon_{0,i}^{\perp}}$ as an initial state to create $\ket{\epsilon_{0,i}}$.
A naive application of the discussion in the previous section would yield an inefficient protocol that scales exponentially with respect to $m$ since the overlap between $\bigotimes_{i=1}^m \ket{\psi_i}$ and $\ket{\epsilon_0}=\bigotimes_{i=1}^m\ket{\epsilon_{0,i}}$ decays as $\prod_i p_i$.
However, instead of directly generating $\bigotimes_{i=1}^m\ket{\epsilon_{0,i}}$ by applying global amplitude amplification to $\bigotimes_{i=1}^m \ket{\psi_i}$, we can independently prepare $\ket{\epsilon_{0,i}}$ using $\log(2/r_i)/\sqrt{p_i}$ applications of filtering polynomial with error parameter $\varepsilon_{\mathrm{filter},i}$.
This replaces $\frac{\log(2/r)}{\sqrt{p}}$ factor with $\sum_i \frac{\log(2/r_i)}{\sqrt{p_i}}$, and does not require the exponential cost.
Note that we should take $r_i=r/m$ and $\varepsilon_{\mathrm{filter},i}=\varepsilon_{\mathrm{filter}}/m$ to maintain sufficient accuracy for $\ket{\epsilon_0}$.
In the following numerical results, we take this approach.

We also perform resource estimation for $m$-acene molecules for $m=4,5,6$.
Geometries of the molecules are taken from Ref. \cite{polyacene}.
We use STO-3G minimal basis set, and obtain an active space consisting of $\pi$-orbitals using PiOS \cite{PiOS}.
The sizes of active spaces are 36, 44, and 52 spin-orbitals for $m=4,5,6$, respectively. 
After the Jordan-Wigner transformation \cite{Jordan1928}, the total Hamiltonian $H_{\mathrm{total}}$ is partitioned into $H$ and $V$ in such a way that, if $\sigma_\ell$ contained in $H_{\mathrm{total}}$ has any Pauli-$X$ or $Y$ operators acting on inactive orbitals, $\sigma_\ell$ is grouped into $V$, and otherwise, $\sigma_\ell$ is taken into $H$.
Note that this choice of perturbation operator $V$ makes the first-order energy correction $\braket{\epsilon_0|V|\epsilon_0}=0$.
We, therefore, do not estimate the cost for first-order perturbation in this case.
Since openfermion \cite{openfermion} is not capable of handling all spin-orbitals of relatively large molecules like polyacene, we derive the expression of molecular Hamiltonians in terms of Majorana operators, each of which corresponds to Pauli operators, in Appendix \ref{appsec:majorana} and use it for calculations.
Since we expect $\epsilon_0^{(2)}\gg\Delta$ in this case, we set $\delta_1=\delta_2 = \delta_2'' = \delta_{\mathrm{chem}}/3$ and $\delta_0 = (\Delta/\epsilon_{\mathrm{MP2}})\delta_2''$, where $\epsilon_{\mathrm{MP2}}$ is the correlation energy obtained with the second-order M{\o}ller-Plesset perturbation theory using restricted Hartree-Fock state as a reference state.

The properties of Hamiltonians are summarized in Table \ref{tab:properties}.
For water clusters, $p$ and $\Delta$ are calculated as the overlap and energy difference between the Hartree-Fock state and the exact ground state of a single water molecule isolated from the other ones.
For polyacene, $\Delta$ is taken from Ref. \cite{polyacene} where the authors calculated the energy gap using density matrix renormalization group.
$p$ is taken to be $0.7$ assuming that we use Hartree-Fock states as input.
This value of $p$ is determined by performing complete-active-space configuration interaction (CASCI) calculation with an active space consisting of 32 spin orbitals and 14 electrons for tetracene and pentacene.
The calculation is performed with OpenMolcas version 22.02 \cite{openmolcas1,openmolcas2} and we find the overlap between the CASCI and Hartree-Fock states are 0.713 and 0.722 respectively for tetracene and pentacene.

Using the numbers in Table \ref{tab:properties}, we calculate the resource according to Sec. \ref{sec:overall-cost} and show the results as Table \ref{tab:first_order_perturbation} and \ref{tab:second_order_perturbation}.
The polynomial degrees are also plotted in Figs.~\ref{fig:filtering} and \ref{fig:ptb}.
The overall cost for the first-order perturbation ranges from the order of $10^{10}$ to $10^{18}$ calls of $U_H$.
That for the second-order is much higher and needs $10^{18}$ to $10^{32}$ calls of $U_H$.
These numbers are not practical; even for the smallest system that we considered, $(\mathrm{H_2O})_2$, the implementation of $U_H$ needs over $10^4$ T gates at least since we find $L_H=3\times 10^3$ (see Sec. \ref{sec:block-encoding}).
This means the overall T-count of the algorithm for the first-order perturbation is over $10^{14}$.
Even if we can implement a T gate in 1 $\mathrm{\mu}$s, the algorithm would take $10^8$ seconds.
Our analysis shows the importance of concrete resource analysis of quantum algorithms beyond $\mathcal{O}$ notation.
The algorithm, at the first sight, seems to be efficient in the sense that it has $\tilde{\mathcal{O}}(\|\bm{h}\|_1\|\bm{v}\|_{2/3}/(\Delta \delta))$ and $\tilde{\mathcal{O}}(\|\bm{h}\|_1\|\bm{v}\|_{2/3}^2/(\Delta^2 \delta))$ cost, but practicality cannot be ensured without such an analysis.

Finally, let us compare this cost with a naive approach that uses phase estimation of $H_{\mathrm{total}}$.
We can well approximate $\mathcal{T}(H_{\mathrm{total}})$ by $\frac{L_H+L_V}{L_H}\mathcal{T}(H) = (1+\frac{L_V}{L_H})\mathcal{T}(H)$ since T-cost is roughly proportional to the number of Pauli terms \cite{Berry2019qubitizationof}.
To obtain the ground state energy with accuracy of $\delta$ by the phase estimation of $H_{\mathrm{total}}$, we need $\sqrt{2}\pi (\|\bm{h}\|_1+\|\bm{v}\|_1)/(2\delta)$ calls of $U_{H_{\mathrm{total}}}$ \cite{PhysRevX.8.041015}.
If a reference state used in phase estimation has overlap $p$ with the true ground state, the correct ground state energy is obtained with probability $p$.
The T-cost for the whole process is therefore well approximated as 
\begin{align}\label{eq:phase-estimation-cost}
    \frac{1}{p}\frac{\sqrt{2}\pi (\|\bm{h}\|_1+\|\bm{v}\|_1)}{2\delta}\left(1+\frac{L_V}{L_H}\right)\mathcal{T}(H)
\end{align}
Taking the numbers from Table \ref{tab:properties}, naive phase estimation of $H_{\mathrm{total}}$ would only need a cost equivalent to $10^{10}$ calls of $U_H$ (calculated roughly through Eq. \eqref{eq:phase-estimation-cost}) even for the largest molecule, hexacene, considered in this work.
Unfortunately, we conclude that the perturbation theory, at least in its present form, does not reduce the computational cost of solving chemical systems.

\section{Conclusion}

In this work, we provided a quantum algorithm to obtain perturbative energies and analyzed its rough computational cost for simple molecular systems.
It is, to the best of our knowledge, the first concrete resource analysis of the practical application of QSP.
Several remarks are in order.

First, the estimated numbers are rather pessimistic; the algorithm needs over $10^{14}$ T gates for the simplest system considered here.
However, it should be remarked that the large contribution to the overall cost comes from the expectation value estimation of the perturbation operator $V$.
This is the same problem faced by the variational quantum eigensolvers \cite{Measurementasaroadbloack} where the energy expectation values have to be determined with high accuracy.
We might be able to reduce the measurement counts, $M^{(1)}$ and $M^{(2)}$, by neglecting unimportant parts of $V$.
For example, when using the active space approximation, we empirically know that energy corrections due to interactions between the core and virtual orbitals are small, and hence might be able to neglect them.
Also, the use of other amplitude estimation techniques such as the ones presented in Ref.~\cite{2207.08628v2} may reduce the runtime by a constant factor.

Second, although the overall cost seems to be impractical, the polynomial degrees are on the order of only $10^8$ even for the largest system we considered.
Using more sophisticated techniques introduced in Ref. \cite{QSP-phases-via-GD} can further reduce the requirement by a factor of 2--3.
Hence, we might be able to perform the generation of the perturbed state (Eq. \eqref{eq:perturbed-state}) in a practical time scale.
The proposed algorithm might therefore become useful when we are interested in measuring very few observables of the perturbed state.
However, it should be noted that we need to obtain QSP phase sequences for $10^8$-degree polynomials.
The state-of-the-art technique for the phase sequence finding \cite{QSP-phases-via-GD} is verified up to $10^4$-degree polynomials but it is not clear if the method still works for such high-order polynomials.

Third, the required numbers for $\varepsilon$ and $r$, which correspond to the infidelity of the prepared states, are very small and become comparable to the gate fidelities of fault-tolerant quantum computers for difficult problems considered in this work.
This is also caused by large $\|V\|$ and small $\delta$, and the requirements are likely to be relaxed for other more simple operators.
However, if we wish to explore the direction considered in this work, the values of $\varepsilon$ and $r$ are likely to remain at the same level.
In this case, gate error rates should also be taken into account for an accurate estimation of required computational resources even in the fault-tolerant quantum computer regime.

Finally, it should be stressed again that the perturbative approach allows us to interpret the physical meaning of the results although the proposed algorithm requires more computational resources for computing total energy than the naive phase estimation approach.
We believe that, while the values of energy are indeed an important quantity, the interpretability of the results is key to the practical applications of quantum simulation algorithms.
This work is only a first step toward this goal, which remains to be reached in the future.

Program code to reproduce the numerical results of this paper is available at \url{https://github.com/kosukemtr/perturbation-resource-estimate}.

\begin{acknowledgments}
    We thank Yasunori Lee, Yuya Nakagawa, and Hideaki Hakoshima for their helpful comments during the finalization of this manuscript.
    KM is supported by JST PRESTO Grant No. JPMJPR2019.
    This work is supported by MEXT Quantum Leap Flagship Program (MEXTQLEAP) Grant No. JPMXS0118067394 and JPMXS0120319794. We also acknowledge support from JST COI-NEXT program Grant No. JPMJPF2014.
    WM is supported by JST PRESTO Grant No. JPMJPR191A
\end{acknowledgments}

\appendix

\section{Polynomial approximation of sign functions and filtering functions}\label{appsec:step-polynomial}

Let
\begin{align}
    \mathrm{sign}(x) = \left\{
        \begin{array}{cl}
            1 & (x>0) \\
            0 & (x=0) \\
            -1 & (x<0)
        \end{array}
    \right.
\end{align}
Low and Chuang \cite{Low2017uniformspectral} first approximates $\mathrm{sign}(x)$ with $\mathrm{erf}(kx)=\frac{2}{\sqrt{\pi}}\int_0^{kx}e^{-y^2} dy$, and then approximates $\mathrm{erf}(kx)$ with a polynomial.
We follow exactly the same strategy and give a detailed cost.

From Lemma 13 and 14 of \cite{Low2017uniformspectral}, we have that, setting
\begin{align}
    n_{\mathrm{exp}}(\beta,\varepsilon)=\lceil\sqrt{2\mathrm{max}\{\beta e^2, \log_e (2/\varepsilon)\}\log(4/\varepsilon)}\rceil,
\end{align}
there exists an $n_{\mathrm{exp}}(\beta,\varepsilon)$-degree polynomial which $\varepsilon$-approximates $e^{-\beta (x+1)}$ for $|x|<1$.
This polynomial can then be utilized to perform an approximation of gaussian $e^{-2\beta x^2}$.
To do this, we simply perform substitution $x\to 2x^2+1$ (see \cite{Low2017uniformspectral}, Corollary 3).
We, therefore, have
\begin{align}
    n_{\mathrm{gauss}}(\beta,\varepsilon)=2\lceil\sqrt{2\mathrm{max}\{\beta e^2, \log_e (2/\varepsilon)\}\log(4/\varepsilon)}\rceil
\end{align}
-degree polynomial suffices for $\varepsilon$-approximation of $e^{-2\beta x^2}$.
\cite{Low2017uniformspectral} then proceeds to approximate $\mathrm{erf}(kx)$ by integrating the polynomial approximation of $e^{-k^2x^2}$ term-by-term.
Equation (71) of \cite{Low2017uniformspectral} shows that, if we use degree-$n$ polynomial which $\varepsilon_{\mathrm{gauss}}$-approximates gaussian, the error in approximating $\mathrm{erf}(kx)$ is,
\begin{align}
    \varepsilon_{\mathrm{erf}}\leq \frac{4k}{(n+1)\sqrt{\pi}} \varepsilon_{\mathrm{gauss}}.
\end{align}

Now, we derive a closed-form expression that shows the degree needed to $\varepsilon$-approximate $\mathrm{erf}(kx)$.
To obtain an analytical solution, we instead find an integer $n$ such that $\varepsilon_{\mathrm{erf}}\leq \frac{4k}{n\sqrt{\pi}} \varepsilon$.
To do this, we want to find the smallest integer $n$ such that, 
\begin{align}
    n&\geq n_{\mathrm{gauss}}\left(\frac{k^2}{2},\frac{n\sqrt{\pi}}{4k}\varepsilon\right)\\
    &= 2\left\lceil\sqrt{2\mathrm{max}\left\{\frac{k^2}{2} e^2, \log_e \left(\frac{8k}{n\sqrt{\pi}\varepsilon}\right)\right\}\log\left(\frac{16k}{n\sqrt{\pi}\varepsilon}\right)}\right\rceil
\end{align}
Note that $\max\left\{k^2 e^2, \log_e \left(\frac{8k}{n\sqrt{\pi}\varepsilon}\right)\right\}$ is almost always equal to $k^2 e^2$ for practical parameter values.
For example, taking $k>10$, $\frac{8k}{n\sqrt{\pi}\varepsilon}$ must be larger than $e^{102}$ to be bigger.
Also, since the ceiling function makes obtaining the analytical solution hard, we instead solve,
\begin{align}
    n &= 2\sqrt{k^2 e^2\log\left(\frac{16k}{n\sqrt{\pi}\varepsilon}\right)}
\end{align}
This equation can be solved analytically, and we conclude that there exists a roughly 
\begin{align}
    n_{\mathrm{erf}}(k,\epsilon) = \frac{16 k}{{\sqrt{\pi} \varepsilon}} \exp\left(- \frac{1}{2}W\left(\frac{128}{\pi \varepsilon^{2} e^{2}}\right)\right)
\end{align}
-degree polynomial which $\varepsilon$-approximates $\mathrm{erf}(kx)$ over $|x|\leq 1$.
Here, $W$ is the Lambert W function.
Let $\tilde{P}^{\mathrm{erf}}_{k,\varepsilon}(x)$ be the polynomial constructed in the above manner.
Then, since $\tilde{P}^{\mathrm{erf}}_{k,\varepsilon}(x)$ approximates $\mathrm{erf}(kx)$ within an error of $\varepsilon$, $|\tilde{P}^{\mathrm{erf}}_{k,\varepsilon}(x)|<1+\varepsilon$.
To implement a polynomial by QSP, it must be bounded by 1.
Here, we define $P^{\mathrm{erf}}_{k,\varepsilon}(x) = \tilde{P}^{\mathrm{erf}}_{k,\varepsilon/2}(x)/(1+\varepsilon/2)$ so that 
\begin{align}
    |P^{\mathrm{erf}}_{k,\varepsilon}(x)| \leq 1
\end{align}
and 
\begin{align}
    |P^{\mathrm{erf}}_{k,\varepsilon}(x) - \mathrm{erf}(kx)|\leq \varepsilon
\end{align}
holds for $|x|\leq 1$.
$P^{\mathrm{erf}}_{k,\varepsilon}(x)$ can be used to construct approximation of shifted error function $\mathrm{erf}(k(x-c))$ for $|c|\leq 1$.
More specifically, we use a polynomial $P^{\mathrm{erf}}_{(1+|c|)k,\varepsilon}\left(\frac{x-c}{1+|c|}\right)$.
It can easily be shown that $\left|P^{\mathrm{erf}}_{(1+|c|)k,\varepsilon}\left(\frac{x-c}{1+|c|}\right)\right|<1$ for $|x|\leq 1$.
The degree required for constructing $P^{\mathrm{erf}}_{(1+|c|)k,\varepsilon}\left(\frac{x-c}{1+|c|}\right)$ is,
\begin{align}
        n_{\mathrm{erf, shifted}}(k,\varepsilon,c) = \frac{32(1+|c|) k}{{\sqrt{\pi} \varepsilon}} \exp\left(- \frac{1}{2}W\left(\frac{512}{\pi \varepsilon^{2} e^{2}}\right)\right).
\end{align}
Note that \cite{Low2017uniformspectral} uses $2$ instead of the factor $1+|c|$, so this is a small improvement over \cite{Low2017uniformspectral}.

Lemma 11 of \cite{Low2017uniformspectral} shows that with 
\begin{align}
    k = \frac{1}{\kappa} \sqrt{2\log_{e}\left(\frac{2}{\pi \varepsilon^2}\right)},
\end{align}
$\mathrm{erf}(kx)$ becomes an $\varepsilon$-approximation of $\mathrm{sign}(x)$ for $|x|>\kappa/2$.
Combining this fact with the above discussion, we conclude that, to $\varepsilon$-approximate $\mathrm{sign}(x-c)$ over the range of $|x-c|>\kappa/2$, the degree of polynomial required can be roughly calculated as,
\begin{align}\label{eq:shifted-sign-degree}
    \begin{split}
        &n_{\mathrm{sign}}(\varepsilon, \kappa, c)\\
        &= \frac{64 (1+|c|)}{\sqrt{\pi} \varepsilon} \frac{1}{\kappa}\sqrt{2\log_e \left(\frac{8}{\pi\varepsilon^2}\right)} \exp\left(- \frac{1}{2}W\left(\frac{2048}{\pi \varepsilon^{2} e^{2}}\right)\right)
    \end{split}
\end{align}
We will hereafter denote the polynomial which $\varepsilon$-approximates $\mathrm{sign}(x-c)$ over the range of $|x-c|>\kappa/2$ by $P^{\mathrm{sign}}_{\varepsilon,\kappa,c}(x)$.

Filtering polynomial used in Sec. \ref{sec:ground-state-preparation} can be readily constructed from $P^{\mathrm{sign}}_{\varepsilon,\kappa,c}(x)$.
Namely, we define 
\begin{align}
    \begin{split}
        &P^{\mathrm{filter}}_{\varepsilon,\kappa,x_{\mathrm{th}}}(x) = \frac{1}{2}\left(P^{\mathrm{sign}}_{\varepsilon,\kappa,-x_{\mathrm{th}}-\kappa/2}(x)-P^{\mathrm{sign}}_{\varepsilon,\kappa,x_{\mathrm{th}}+\kappa/2}(x)\right).
    \end{split}
\end{align}
It is easy to verify its properties \eqref{eq:filtering-property} and degree \eqref{eq:filtering-degree}.

\section{Error analysis for the state preparation}\label{appsec:prep-error}
By Eq. \eqref{eq:filtering-property},
\begin{align}
    P^{\mathrm{filter}}_{\varepsilon,\kappa,x_{\mathrm{th}}}\left(\frac{H'}{\|\bm{h}'\|_1}\right)\ket{\psi} &= (1-\varepsilon)\sqrt{p}\ket{\epsilon_0} + \varepsilon\sqrt{1-p}\ket{\epsilon_0^\perp}.
\end{align}
Therefore,
\begin{align}
    \begin{split}
        &\bra{\psi}P^{\mathrm{filter}}_{\varepsilon,\kappa,x_{\mathrm{th}}}\left(\frac{H'}{\|\bm{h}'\|_1}\right)O P^{\mathrm{filter}}_{\varepsilon,\kappa,x_{\mathrm{th}}}\left(\frac{H'}{\|\bm{h}'\|_1}\right)\ket{\psi}\\
        &= (1-2\varepsilon)p\braket{\epsilon_0|O|\epsilon_0} + 2\varepsilon\sqrt{p(1-p)}\mathrm{Re}\braket{\epsilon_0^{\perp}|O|\epsilon_0} + \mathcal{O}(\varepsilon^2).    
    \end{split}
    \label{eq:appendixB1}
\end{align}
Now, the norm can be approximated as,
\begin{align}
    \left\|P^{\mathrm{filter}}_{\varepsilon,\kappa,x_{\mathrm{th}}}\left(\frac{H'}{\|\bm{h}'\|_1}\right)\ket{\psi}\right\|^2 &= p(1 - 2\varepsilon) + \mathcal{O}(\varepsilon^2), \label{appendixB2}
\end{align}
Using Eqs. \eqref{eq:appendixB1} and \eqref{appendixB2}, we conclude,
\begin{align}
    \braket{\tilde{\epsilon}_0|O|\tilde{\epsilon}_0} &= \braket{\epsilon_0|O|\epsilon_0} + 2\varepsilon\sqrt{\frac{1-p}{p}}\mathrm{Re}\braket{\epsilon_0^{\perp}|O|\epsilon_0} + \mathcal{O}(\varepsilon^2),
\end{align}
which leads to Eq. \eqref{eq:delta_prep}.

\section{Optimal cost of robust amplitude estimation}\label{appsec:optimal-dist-RAE}
We consider an optimal distribution of $M_\ell$ to achieve an overall mean squared error $\delta^2$ in estimating $\braket{\tilde{\tilde{\epsilon}}_0|V|\tilde{\tilde{\epsilon}}_0}=\sum_\ell v_\ell \braket{\tilde{\tilde{\epsilon}}_0|\sigma_\ell|\tilde{\tilde{\epsilon}}_0}$.
For a given distribution of $M_\ell$, the overall error $\delta^2_{\{M_\ell\}}$ can be written as,
\begin{align}
    \delta^2_{\{M_\ell\}} = \sum_\ell v_\ell^2 \delta_\ell^2,
\end{align}
since estimates of $\braket{\tilde{\tilde{\epsilon}}_0|\sigma_\ell|\tilde{\tilde{\epsilon}}_0}$ are independent from each other.
We wish to minimize $M=\sum_\ell M_\ell$ while achieving $\delta^2_{\{M_\ell\}}=\delta^2$.
To do this, we define a Lagrangian
\begin{align}
    \mathcal{L} = \sum_\ell M_\ell + \lambda\left(\sum_\ell v_\ell^2 \delta_\ell^2-\delta^2\right),
\end{align}
and solve $\frac{\partial \mathcal{L}}{\partial M_\ell}=0$.
This leads to,
\begin{align}
    M_\ell = (2\alpha^2 v_\ell^2 \lambda)^{1/3}, 
\end{align}
where $\alpha = \frac{5\sqrt{2}}{2}\frac{e^2}{e-1}$.
Plugging this into the constraint $\delta^2_{\{M_\ell\}}=\delta^2$, we get,
\begin{align}
    \delta^2 = \sum_\ell \alpha^2 v_\ell^2 \frac{1}{(2\alpha^2 v_\ell^2 \lambda)^{2/3}},
\end{align}
which is equivalent to
\begin{align}
    \lambda^{1/3} = \frac{\alpha^{1/3}}{2^{1/3}\delta}\sqrt{\sum_\ell v_\ell^{2/3}} 
\end{align}
We, therefore, conclude the optimal distribution of $M_\ell$ is
\begin{align}
    M_\ell = \frac{\alpha v_\ell^{2/3}}{\delta}\sqrt{\sum_\ell v_\ell^{2/3}}, 
\end{align}
and total calls $M$ of $W$ is,
\begin{align}
    M = \sum_\ell M_\ell =  \frac{5\sqrt{2}}{2}\frac{e^2}{e-1}\frac{1}{\delta}\left(\sum_\ell v_\ell^{2/3}\right)^{3/2}.
\end{align}

\section{Polynomial for perturbation} \label{appsec:perturbation}
First, we review the method presented in Ref. \cite{GrandUnification} to construct a QSP-implementable polynomial that approximates $1/x$.
Let 
\begin{align}
    P^{\mathrm{inv}}_{\varepsilon,w}(x) &= \sum_{j=0}^{D(\varepsilon, w)}(-1)^{j}c_j T_{2 j+1}(x)  
\end{align}
where $T_{i}(x)$ is the Chebyshev polynomial of order $i$, and
\begin{align}
    D(\varepsilon, w) &=\left\lceil\sqrt{b(\varepsilon, w) \log (4 b(\varepsilon, w) / \varepsilon)}\right\rceil, \label{eq:D} \\
    c_j &= 4\times 2^{-2 b} \sum_{i=j+1}^{b}\left(\begin{array}{c}
        2 b \\
        b+i
        \end{array}\right) \label{eq:c}
\end{align}
where 
\begin{align}
    b(\varepsilon, w) = \left\lceil\left(\frac{1}{w}\right)^{2} \log \left(\frac{1}{w\varepsilon}\right)\right\rceil
\end{align}
This function $P^{\mathrm{inv}}_{\varepsilon,w}(x)$ is a $2\varepsilon$-approximation of $1/x$ for $w<|x|<1$ \cite{Childs2017QLSP}.
Moreover, it is notable that from the function form of $f_\varepsilon(x)$, it holds that, 
\begin{align}
    \left|P^{\mathrm{inv}}_{\varepsilon,w}(x)\right|
    &\leq |x|\left.\frac{\partial f_\varepsilon}{\partial x}\right|_{x=0}\\
    &\leq |x|\sum_{j=0}^D c_j \left|\frac{\partial T_{2j+1}(0)}{\partial x}\right| \\
    &\leq 4|x|\sum_{j=0}^D (2j+1) \\
    &\leq 4|x|(D+1)^2, \label{eq:inv-maximum}
\end{align}
for any $x$, since the coefficient $c_j$ satisfies $c_j \leq 4$ and $\left|\frac{\partial T_n(0)}{\partial x}\right| = n$.

Ref. \cite{GrandUnification} multiplies a polynomial defined as,
\begin{align}
    \begin{split}
        &P_{\varepsilon, w}^{\mathrm{rect}}(x):=\\
        & \frac{1}{1+\frac{\varepsilon}{2}}\left\{1+\frac{1}{2}\left[P_{\varepsilon, \frac{w}{4}, \frac{3w}{4}}^{\mathrm{sign}}\left(x\right)-P_{\varepsilon, \frac{w}{4}, -\frac{3w}{4}}^{\mathrm{sign}}\left(x\right)\right]\right\}
    \end{split}
\end{align}
to $P_{\varepsilon,w}^{\mathrm{inv}}(x)$ to make a QSP-implementable polynomial that $\epsilon$-approximates $w/(2x)$.
Note that $P_{\varepsilon, w}^{\mathrm{rect}}(x)$ satisfies
\begin{align}
    |P_{\varepsilon, w}^{\mathrm{rect}}(x)| &> 1-\varepsilon ~~(w<|x|<1)\\
    |P_{\varepsilon, w}^{\mathrm{rect}}(x)| &< \varepsilon ~~(|x|<\frac{w}{2}) \label{eq:Prect_zero}
\end{align}
More concretely, we define \textit{matrix inversion polynomial} \cite{GrandUnification} as
\begin{align}
    P^{\mathrm{MI}}_{\varepsilon, w}(x) = \frac{w}{2}P^{\mathrm{inv}}_{\varepsilon/2, w/2}(x)P^{\mathrm{rect}}_{\varepsilon', w}(x),
\end{align}
where
\begin{align}
    \varepsilon' = \min\left\{\frac{2\varepsilon w}{5},~\frac{1}{2wD(\frac{\varepsilon}{4},\frac{w}{2})}\right\}.
\end{align}
This function satisfies
\begin{align}
    \left|P^{\mathrm{MI}}_{\varepsilon, w}(x)-\frac{w}{2}\frac{1}{x}\right|<\frac{w}{2}\varepsilon
\end{align}
for $|x|>w$. Note that from the construction, the degree of the polynomial is,
\begin{align}
    n_{\mathrm{MI}} &= 2D\left(\frac{\varepsilon}{4}, \frac{w}{2}\right) + n_{\mathrm{sign}}\left(\varepsilon', \frac{w}{4}, \frac{3w}{4}\right).
\end{align}

$P^{\mathrm{MI}}_{\varepsilon, w}(x)$ is constructed for matrix inversion, but it can also be used for perturbation.
It is because of the property of $P_{\varepsilon, w}^{\mathrm{rect}}(x)$ that it becomes almost zero around $x=0$ (Eq. \eqref{eq:Prect_zero}).
$P^{\mathrm{MI}}_{\varepsilon, w}\left(\frac{H-\epsilon_0}{\|\bm{h}'\|_1}\right)$ hence approximates $\|\bm{h}'\|_1\Pi (H-\epsilon_0)^{-1} \Pi$.
However, $P^{\mathrm{MI}}_{\varepsilon, w}(x)$ does not control the error at around $x=0$;
from Eqs. \eqref{eq:inv-maximum} and \eqref{eq:Prect_zero}, we can only guarantee that 
\begin{align}
    \left|P^{\mathrm{MI}}_{\varepsilon, w}(x)\right|< 2w|x|(D+1)^2\varepsilon'
\end{align}
for $|x|<w/2$.
We, therefore, introduce another parameter $w_0$ and define $\varepsilon''$ as
\begin{align}
    \varepsilon'' = \min\left\{\frac{2\varepsilon w}{5},~\frac{1}{4wD(\frac{\varepsilon}{4},\frac{w}{2})},~\frac{\varepsilon}{2w_0 [D(\frac{\varepsilon}{4},\frac{w}{2})+1]^2}\right\}.
\end{align}
Using $\varepsilon''$, let
\begin{align}
    P^{\mathrm{ptb}}_{\varepsilon, w, w_0}(x) := \frac{w}{2}P^{\mathrm{inv}}_{\varepsilon/2, w/2}(x)P^{\mathrm{rect}}_{\varepsilon'', w}(x).
\end{align}
Then, it is easy to see Eqs. \eqref{eq:Pptb_inv} and \eqref{eq:Pptb_zero} holds.
Its degree is
\begin{align}
    n_{\mathrm{ptb}}(\varepsilon, w, w_0) &= 2D\left(\frac{\varepsilon}{4}, \frac{w}{2}\right) + n_{\mathrm{sign}}\left(\varepsilon'', \frac{w}{4}, \frac{3w}{4}\right)
\end{align}

\section{Error in second-order perturbation energy}\label{appsec:second-order-error}
First, note that,
\begin{align}
    \tilde{\epsilon}_0^{(2)} &= -\frac{2}{w\|\bm{h}'\|_1}\bra{\epsilon_0}VP^{\mathrm{ptb}}_{\varepsilon, w, w_0}\left(\frac{H-\hat{\epsilon}_0}{\|\bm{h}'\|_1}\right)V\ket{\epsilon_0}
\end{align}
Let $V_{ij} = \bra{\epsilon_i}V\ket{\epsilon_j}$. Then, the above can be expanded as,
\begin{align}
    \begin{split}
        \tilde{\epsilon}_0^{(2)}
        &=  -\frac{2}{w\|\bm{h}'\|_1}|V_{00}|^2 P^{\mathrm{ptb}}_{\varepsilon, w, w_0}\left(\frac{\epsilon_0-\hat{\epsilon_0}}{\|\bm{h}'\|_1}\right) \\
        &\quad-  \frac{2}{w\|\bm{h}'\|_1}\sum_{i\neq 0}|V_{i0}|^2 P^{\mathrm{ptb}}_{\varepsilon, w, w_0}\left(\frac{\epsilon_i-\hat{\epsilon_0}}{\|\bm{h}'\|_1}\right).    
    \end{split}
    \label{eq:braketP-decomposed}
\end{align}
Note that the second term approximates $\epsilon_0^{(2)}$.
Now, by the assumption of Eqs. \eqref{eq:w-concrete} and \eqref{eq:w0-concrete}, 
\begin{align}
    |P^{\mathrm{ptb}}_{\varepsilon, w, w_0}(x)-w/(2x)|&\leq \frac{w}{2}\varepsilon ~~\left(\frac{\Delta-\delta_0}{\|\bm{h}'\|_1}<|x|<1\right),\\
    |P^{\mathrm{ptb}}_{\varepsilon, w, w_0}(x)|&\leq \frac{w}{2}\varepsilon ~~\left(|x|<\frac{\delta_0}{\|\bm{h}'\|_1}\right).
\end{align}
Then, it holds that,
\begin{align}
    \left|\frac{2}{w\|\bm{h}'\|_1}P^{\mathrm{ptb}}_{\varepsilon, w, w_0}\left(\frac{\epsilon_i-\hat{\epsilon_0}}{\|\bm{h}'\|_1}\right) - \frac{1}{\epsilon_i - \hat{\epsilon}_0} \right|&\leq \frac{\varepsilon}{\|\bm{h}'\|_1}\\
    \left|\frac{2}{w\|\bm{h}'\|_1}P^{\mathrm{ptb}}_{\varepsilon, w, w_0}\left(\frac{\epsilon_0-\hat{\epsilon_0}}{\|\bm{h}'\|_1}\right)\right| &\leq \frac{\varepsilon}{\|\bm{h}'\|_1}
\end{align}
for all $\epsilon_i$.
Also, we have,
\begin{align}
    \left|\frac{1}{\epsilon_i-\hat{\epsilon_0}}-\frac{1}{\epsilon_i-\epsilon_0}\right| \leq \frac{\delta_0}{(\epsilon_i-\epsilon_0) (\epsilon_i-\epsilon_0-\delta_0)}.
\end{align}
Therefore,
\begin{align}
    \begin{split}
        &\left|\frac{2}{w\|\bm{h}'\|_1}P^{\mathrm{ptb}}_{\varepsilon, w, w_0}\left(\frac{\epsilon_i-\hat{\epsilon_0}}{\|\bm{h}'\|_1}\right) - \frac{1}{\epsilon_i - \epsilon_0} \right|\\
        &\leq \frac{\varepsilon}{\|\bm{h}'\|_1} + \frac{\delta_0}{(\epsilon_i-\epsilon_0) (\epsilon_i-\epsilon_0-\delta_0)}.
    \end{split}
\end{align}
Noting that $\epsilon^{(2)}_0$ can also be written as
\begin{align}
    \epsilon^{(2)}_0 = -\sum_{i\neq 0}\frac{|V_{i0}|^2}{\epsilon_i-\epsilon_0},
\end{align}
we find,
\begin{align}
    &\left|\epsilon_{0}^{(2)} - \tilde{\epsilon}_{0}^{(2)}\right| \nonumber\\
    &\leq \frac{\varepsilon|V_{00}|^2}{\|\bm{h}'\|_1}+\sum_{i\neq 0}|V_{i0}|^2 \left(\frac{\varepsilon}{\|\bm{h}'\|_1} + \frac{\delta_0}{(\epsilon_i-\epsilon_0) (\epsilon_i-\epsilon_0-\delta_0)}\right)\\
    &\leq \frac{\|V\ket{\epsilon_0}\|^2 }{\|\bm{h}'\|_1}\varepsilon + \sum_{i\neq 0}\frac{|V_{i0}|^2}{\epsilon_i-\epsilon_0} \frac{\delta_0\|\bm{h}'\|_1}{(\Delta-\delta_0)} \\
    &= \frac{\|V\ket{\epsilon_0}\|^2 }{\|\bm{h}'\|_1}\varepsilon + \frac{\delta_0}{(\Delta-\delta_0)}\epsilon_{0}^{(2)}. 
    \label{eq:error-ineq0}
\end{align}

\section{Writing the molecular Hamiltonian in terms of Majorana operators}\label{appsec:majorana}

The electronic structure of molecules is determined by the following Hamiltonian:
\begin{align}
    \mathcal{H}=\sum_{p q} \sum_\sigma h_{p q} a_{p \sigma}^{\dagger} a_{q \sigma}+\frac{1}{2} \sum_{p q r s} \sum_{\sigma \tau} g_{p q r s} a_{p \sigma}^{\dagger} a_{r \tau}^{\dagger} a_{s \tau} a_{q \sigma},
\end{align}
where $a_{p\sigma}^\dagger$ and $a_{p\sigma}$ are creation and annihilation operators, respectively.
This Hamiltonian can be expressed in terms of Majorana operators defined as
\begin{align}
    \gamma_{p \sigma, 0}=a_{p \sigma}+a_{p \sigma}^{\dagger}, \quad \gamma_{p \sigma, 1}=-i\left(a_{p \sigma}-a_{p \sigma}^{\dagger}\right),
\end{align}
as,
\begin{align}
    \begin{split}
        \mathcal{H}=&\left(\sum_p h_{p p}+\frac{1}{2} \sum_{p r} g_{p p r r}-\frac{1}{4} \sum_{p r} g_{p r r p}\right) \mathcal{I} \\
    &+\frac{i}{2} \sum_{p q \sigma}\left(h_{p q}+\sum_r g_{p q r r}-\frac{1}{2} \sum_r g_{p r r q}\right) \gamma_{p \sigma, 0} \gamma_{q \sigma, 1} \\
    &+\frac{1}{4} \sum_{p>r, s>q} \sum_\sigma\left(g_{p q r s}-g_{p s r q}\right) \gamma_{p \sigma, 0} \gamma_{r \sigma, 0} \gamma_{q \sigma, 1} \gamma_{s \sigma, 1} \\
    &+\frac{1}{8} \sum_{p q r s} \sum_{\sigma \neq \tau} g_{p q r s} \gamma_{p \sigma, 0} \gamma_{r \tau, 0} \gamma_{q \sigma, 1} \gamma_{s \tau, 1},
    \end{split}\label{appeq:majorana_hamiltonian_previous}
\end{align}
which is derived in \cite{OR_1norm_reduction}.
We can further simplify the last summation of the above expression by employing the symmetry of operators and $g_{pqrs}$.
First, let us define
\begin{align}
    A_{prqs} = \sum_{\sigma \neq \tau} \gamma_{p \sigma, 0} \gamma_{r \tau, 0} \gamma_{q \sigma, 1} \gamma_{s \tau, 1}.
\end{align}
Then, the last summation in Eq. \eqref{appeq:majorana_hamiltonian_previous} can be written as,
\begin{align}
    \begin{split}
        &\sum_{p q r s} \sum_{\sigma \neq \tau} g_{p q r s} \gamma_{p \sigma, 0} \gamma_{r \tau, 0} \gamma_{q \sigma, 1} \gamma_{s \tau, 1} = \sum_{p q r s}  g_{p q r s} A_{prqs}.
    \end{split}
\end{align}
This can be simplified using the symmetry of $A_{prqs}$.
$A_{prqs}$ has symmetry about permutations $p\leftrightarrow r$ and $q \leftrightarrow s$.
Namely,
\begin{align}
    A_{prqs} = A_{rpqs}.
\end{align}
Using the above property and the symmetry of $g_{pqrs}$, we can simplify the expression as follows:
\begin{align}
    \begin{split}
        \sum_{p q r s}  g_{p q r s} A_{prqs} = &2\sum_{p>r}\sum_{q\neq s} g_{pqrs} A_{prqs} \\ 
        &+ 2\sum_q \sum_{p>r} g_{pqrq} A_{prqq} \\
        &+ 2\sum_p \sum_{q>s} g_{pqps} A_{ppqs} \\ 
        &+ \sum_{p,q} g_{pqpq} A_{ppqq} \\
    \end{split}
\end{align}
Substituting the definition of $A_{prqs}$ back, we get,
\begin{align}
    \begin{split}
        &\sum_{p q r s} \sum_{\sigma \neq \tau} g_{p q r s} \gamma_{p \sigma, 0} \gamma_{r \tau, 0} \gamma_{q \sigma, 1} \gamma_{s \tau, 1} = \\
        & 2\sum_{p> r}\sum_{q\neq s}\sum_{\sigma\neq \tau} g_{pqrs} \gamma_{p \sigma, 0} \gamma_{r \tau, 0} \gamma_{q \sigma, 1} \gamma_{s \tau, 1} \\
        &+ 2\sum_q \sum_{p>r} \sum_{\sigma\neq \tau} g_{pqrq} \gamma_{p \sigma, 0} \gamma_{r \tau, 0} \gamma_{q \sigma, 1} \gamma_{q \tau, 1} \\
        &+ 2\sum_p \sum_{q>s}\sum_{\sigma\neq \tau} g_{pqps} \gamma_{p \sigma, 0} \gamma_{p \tau, 0} \gamma_{q \sigma, 1} \gamma_{s \tau, 1} \\
        &+ 2\sum_{p,q} g_{pqpq} \gamma_{p \alpha, 0} \gamma_{p \beta, 0} \gamma_{q \alpha, 1} \gamma_{q \beta, 1}.
    \end{split}
\end{align}
Note that we used $A_{ppqq} = 2\gamma_{p \alpha, 0} \gamma_{p \beta, 0} \gamma_{q \alpha, 1} \gamma_{q \beta, 1}$.
Finally, we obtain,
\begin{align}
    \begin{split}
        \mathcal{H}=&\left(\sum_p h_{p p}+\frac{1}{2} \sum_{p r} g_{p p r r}-\frac{1}{4} \sum_{p r} g_{p r r p}\right) \mathcal{I} \\
    &+\frac{i}{2} \sum_{p q \sigma}\left(h_{p q}+\sum_r g_{p q r r}-\frac{1}{2} \sum_r g_{p r r q}\right) \gamma_{p \sigma, 0} \gamma_{q \sigma, 1} \\
    &+\frac{1}{4} \sum_{p>r, s>q} \sum_{\sigma}\left(g_{p q r s}-g_{p s r q}\right) \gamma_{p \sigma, 0} \gamma_{r \tau, 0} \gamma_{q \sigma, 1} \gamma_{s \tau, 1} \\
    &+ \frac{1}{4}\sum_{p> r}\sum_{q\leq s}\sum_{\sigma\neq \tau} g_{pqrs} \gamma_{p \sigma, 0} \gamma_{r \tau, 0} \gamma_{q \sigma, 1} \gamma_{s \tau, 1} \\
    &+ \frac{1}{4}\sum_{p\geq r} \sum_{q>s}\sum_{\sigma\neq \tau} g_{pqrs} \gamma_{p \sigma, 0} \gamma_{r \tau, 0} \gamma_{q \sigma, 1} \gamma_{s \tau, 1} \\
    &+ \frac{1}{4}\sum_{p,q} g_{pqpq} \gamma_{p \alpha, 0} \gamma_{p \beta, 0} \gamma_{q \alpha, 1} \gamma_{q \beta, 1}.
    \end{split}
\end{align}

\bibliographystyle{apsrev4-2}
\bibliography{bib}

\end{document}